\documentclass[11pt]{article}
\pdfoutput=1
\usepackage{amsmath,amsfonts,amssymb}
\usepackage{cite,url}
\usepackage{epsfig}
\usepackage{graphicx}
\usepackage[update,prepend]{epstopdf}
\usepackage[utf8]{inputenc} 
\usepackage[bottom]{footmisc}

\allowdisplaybreaks

\def\smallSM{{\rm{\scriptscriptstyle SM}}}
\def\smallMSSM{{\rm{\scriptscriptstyle MSSM}}}
\def\smallFSSM{{\rm{\scriptscriptstyle FSSM}}}

\def\smallZ{{\scriptscriptstyle Z}}
\def\smallW{{\scriptscriptstyle W}}

\def\smallH{{\scriptscriptstyle H}}

\def\smallL{{\scriptscriptstyle L}}
\def\smallR{{\scriptscriptstyle R}}

\def\MS{M_S}
\def\MZ{M_\smallZ}
\def\MW{M_\smallW}

\def\beq{\begin{equation}}
\def\eeq{\end{equation}}
\def\bea{\begin{eqnarray}}
\def\eea{\end{eqnarray}}
\def\nn{\nonumber}
\def\wt{\widetilde}

\def\sq2{\sqrt{2}}
\def\drbar{{\ensuremath{ \overline{\rm DR}}}}
\def\msbar{\overline{\rm MS}}

\def\qew{Q_{{\rm{\scriptscriptstyle EW}}}}

\def\gm2{(g-2)_\mu}
\newcommand{\lambdaSM}{\lambda_{\smallSM}}
\def\amuexp{a_\mu^{\rm exp}}
\def\amuSM{a_\mu^{\smallSM}}
\def\damu{\Delta a_\mu}

\def\gp{g^\prime}
\def\gpq{g^{\prime\,2}}
\def\gpqq{g^{\prime\,4}}

\long\def\symbolfootnote[#1]#2{\begingroup%
\def\thefootnote{\fnsymbol{footnote}}\footnote[#1]{#2}\endgroup}

\newenvironment{Appendix}
 {
  \setcounter{section}{0}
  \setcounter{equation}{0}
  
 }{}

\makeatletter
\newcommand{\vast}{\bBigg@{3}}
\makeatother

\begin{document}

\begin{titlepage}

%\begin{flushright}
%\today
%\end{flushright}

\mbox{}
\vspace{1.5cm}
\begin{center}

\vspace{1cm}

{\LARGE \bf Higgs-mass constraints on a supersymmetric solution} 
\vskip 0.3cm
{\LARGE \bf of the muon $g-2$ anomaly}

\vspace{1cm}

{\Large Wenqi Ke$^{\,a}$ and Pietro~Slavich$^{\,a}$}

\vspace*{5mm}

{\sl ${}^a$ Sorbonne Université, CNRS, Laboratoire de Physique
  Th\'eorique et Hautes Energies (LPTHE), UMR 7589, 4 place Jussieu,
  75005 Paris, France.}
\end{center}
\symbolfootnote[0]{{\tt e-mail:}}
\symbolfootnote[0]{{\tt wke@lpthe.jussieu.fr}}
\symbolfootnote[0]{{\tt slavich@lpthe.jussieu.fr}}

\vspace{0.7cm}

\abstract{The prediction for the quartic coupling of the SM-like Higgs
  boson constrains the parameter space of SUSY models, even in
  scenarios where all of the new-particle masses are above the scale
  probed so far by the LHC. We study the implications of the
  Higgs-mass prediction on a recently-proposed SUSY model that
  features two pairs of Higgs doublets, and provides a solution to the
  $\gm2$ anomaly thanks to a suitable enhancement of the muon Yukawa
  coupling.}

\vfill

\end{titlepage}

%\tableofcontents

\setcounter{footnote}{0}

\section{Introduction}
\label{sec:intro}

The discovery of a Higgs boson with mass around $125$~GeV and
properties compatible with the predictions of the Standard Model
(SM)~\cite{CMS:2012qbp, ATLAS:2012yve, ATLAS:2015yey, ATLAS:2016neq},
combined with the negative (so far) results of the searches for
additional new particles at the LHC, point to scenarios with at least
a mild hierarchy between the electroweak (EW) scale and the scale of
beyond-the-SM (BSM) physics. In this case, the SM plays the role of an
effective field theory (EFT) valid between the two scales. The
requirement that a given BSM model include a a state that can be
identified with the observed Higgs boson can translate into important
constraints on the model's parameter space.

One of the prime candidates for BSM physics is supersymmetry (SUSY),
which predicts scalar partners for all SM fermions, as well as
fermionic partners for all bosons. A remarkable feature of SUSY
extensions of the SM is the requirement of an extended Higgs sector,
with additional neutral and charged bosons. In contrast to the case of
the SM, the masses of the Higgs bosons are not free parameters, as
SUSY requires all quartic scalar couplings to be related to the gauge
and Yukawa couplings. Moreover, radiative corrections to the
tree-level predictions for the quartic scalar couplings introduce a
dependence on all of the SUSY-particle masses and
couplings.\footnote{We point the reader to ref.~\cite{Slavich:2020zjv}
for a recent review of Higgs-mass predictions in SUSY models.} In a
hierarchical scenario such as the one described above, the prediction
of the SUSY model for the quartic self-coupling of its lightest Higgs
scalar, which plays the role of the SM Higgs boson, must coincide with
the SM coupling $\lambdaSM$ extracted at the EW scale from the
measured value of the Higgs mass and evolved up to the SUSY scale with
appropriate renormalization group equations (RGEs). This condition can
be used to constrain some yet-unmeasured parameters of the SUSY model,
such as, e.g., the masses of the scalar partners of the top quarks,
the stops.

While the new particles predicted by SUSY models -- or, for that
matter, those predicted by any other BSM model -- have yet to show up
at the LHC, precision experiments have seen tantalizing deviations
from the predictions of the SM, particularly in measurements involving
muons. Over the past few years the LHCb collaboration reported hints
of lepton flavor violation in rare $B$ decays~\cite{LHCb:2014vgu,
  LHCb:2017avl, LHCb:2019hip, LHCb:2021trn}, and earlier in 2021 the
Muon g-2 Collaboration at Fermilab reported a new
measurement~\cite{Muong-2:2021ojo} of the muon anomalous magnetic
moment $a_\mu \equiv \gm2/2$, consistent with the previous measurement
by the E821 experiment at BNL~\cite{Muong-2:2006rrc}. In what might be
considered currently the most striking deviation from the predictions
of the SM, the combination of the two experimental results for the
muon anomalous magnetic moment, $\amuexp =
116\,592\,061(41)\,\times\,10^{-11}$, differs by $4.2\,\sigma$ from
the state-of-the-art SM prediction given in
ref.~\cite{Aoyama:2020ynm}, $\amuSM =
116\,591\,810(43)\,\times\,10^{-11}$, which is based on
refs.~\cite{Aoyama:2012wk, Aoyama:2019ryr, Czarnecki:2002nt,
  Gnendiger:2013pva, Davier:2017zfy, Keshavarzi:2018mgv,
  Colangelo:2018mtw, Hoferichter:2019gzf, Davier:2019can,
  Keshavarzi:2019abf, Kurz:2014wya, Melnikov:2003xd, Masjuan:2017tvw,
  Colangelo:2017fiz, Hoferichter:2018kwz, Gerardin:2019vio,
  Bijnens:2019ghy, Colangelo:2019uex, Blum:2019ugy,
  Colangelo:2014qya}.

Supersymmetric extensions of the SM can accommodate an explanation for
the observed discrepancy $\damu \equiv \amuexp-\amuSM = (251 \pm
59)\,\times\,10^{-11}$.  In the minimal of such extensions, the MSSM,
a suitable contribution to $a_\mu$ can arise from one-loop diagrams
involving smuons, higgsinos and EW gauginos (namely, the SUSY partners
of muons, Higgs bosons and EW gauge bosons). This contribution is
suppressed by the ratio $M_\mu^2/\MS^2$ -- where $M_\mu$ is the muon
mass and $\MS$ represents the mass scale of the relevant SUSY
particles -- but it can be enhanced by a large value of the parameter
$\tan\beta \equiv v_u/v_d$, i.e.~the ratio of the vacuum expectation
values (vevs) of the Higgs doublets $H_u$ and $H_d$, which give mass
to the up-type and down-type fermions, respectively. However, since
the Yukawa couplings of the down-type fermions $f_d$ in the MSSM are
related to their SM counterparts by $y^\smallMSSM_{f_d} =
g^\smallSM_{f_d}/\cos\beta$, the requirement that the bottom and tau
couplings remain perturbative up to the GUT scale sets an upper limit
on the acceptable values of $\tan\beta$, see
e.g.~ref.~\cite{Altmannshofer:2010zt}. When such limit is taken into
account, the masses of the SUSY particles entering the diagrams that
provide the required contribution to $a_\mu$ are typically restricted
to the few-hundred-GeV range. This results in some tension with the
direct searches for SUSY particles at the LHC, although specific
regions of the MSSM parameter space -- typically, those with a
``compressed'' SUSY mass spectrum -- remain still viable.\footnote{For
recent surveys of explanations of the $\gm2$ anomaly in the MSSM see
e.g.~refs.~\cite{Chakraborti:2021dli, Athron:2021iuf}. For an earlier
study of $\gm2$ in MSSM scenarios with TeV-scale SUSY masses and very
large $\tan\beta$ see ref.~\cite{Bach:2015doa}.}

Recently, a new SUSY model in which a suitable contribution to $a_\mu$
can be obtained even with smuon, higgsino and gaugino masses in the
multi-TeV range was proposed in ref.~\cite{Altmannshofer:2021hfu}. The
Higgs sector of the ``Flavorful Supersymmetric Standard Model''
(FSSM)\footnote{We remark that this acronym had already been used in
  ref.~\cite{Benakli:2013msa} to denote a model with ``Fake'' Split
  SUSY.}  consists of four doublets, two of which, $H_u$ and $H_d$,
couple only to quarks and leptons of the third generation, whereas the
other two, $H_u^\prime$ and $H_d^\prime$, have much smaller vevs and
provide masses to the fermions of the first and second generation. In
this model the muon Yukawa coupling $y^\smallFSSM_{\mu}$, which
determines the higgsino--muon--smuon and Higgs--smuon--smuon couplings
entering the diagrams that contribute to $a_\mu$, can be of ${\cal
  O}(1)$ without implying non-perturbative values for
$y^\smallFSSM_{\tau}$ and $y^\smallFSSM_{b}$. Indeed, the SUSY
contribution to $a_\mu$ in the FSSM is enhanced by $v_u/v_d^\prime$,
which can greatly exceed the enhancement achievable in the MSSM when
$v_d^\prime \ll v_d$, in turn allowing for a stronger suppression by
$M_\mu^2/\MS^2$.

Scenarios where all of the SUSY particles have masses in the multi-TeV
range will be probed directly only at future colliders. However, as
discussed in ref.~\cite{Altmannshofer:2021hfu}, the extended
Higgs/higgsino sector of the FSSM can accommodate interesting
flavor-changing effects both in the lepton sector and in the quark
sector, leading to constraints on the flavor structure of the Yukawa
couplings. As mentioned earlier, a further constraint stems from the
requirement that the lightest scalar in the Higgs sector be identified
with the SM-like Higgs boson discovered at the LHC.  Compared with the
case of the MSSM, the presence of additional particles in the
Higgs/higgsino sector and of additional ${\cal O}(1)$ couplings in the
superpotential can affect the FSSM prediction for the SM-like Higgs
mass, leading to different constraints on the parameter space of the
model.

\bigskip

In this paper we study the Higgs-mass prediction of the FSSM and its
interplay with the solution of the $\gm2$ anomaly. In the calculation
of the Higgs mass we rely on the EFT approach, as appropriate to a
hierarchical scenario where the BSM physics is somewhat removed from
the EW scale. In section~\ref{sec:spectrum} we introduce the Higgs
sector of the FSSM. In section~\ref{sec:matching} we obtain the
one-loop threshold correction to the quartic Higgs coupling, adapting
to the model under consideration the general formulas given in
ref.~\cite{Braathen:2018htl}. Combined with two-loop RGEs for the SM
couplings, this allows for the next-to-leading-logarithmic (NLL)
resummation of the corrections to the Higgs mass enhanced by powers of
$\ln(\MS/M_t)$ (where, as usual, we take $M_t$ as a proxy for the EW
scale).  We also point out a potential issue stemming from large
threshold corrections to the strange Yukawa coupling in case the
four-doublet construction of the FSSM is extended to the quark
sector. In section~\ref{sec:numerical} we discuss the constraints on
the parameter space of the FSSM that arise from the combined
requirements of an appropriate prediction for the Higgs mass and a
solution to the $\gm2$ anomaly. Section~\ref{sec:conclusions} contains
our conclusions. Finally, in the appendix we provide explicit formulas
for the tree-level Higgs mass matrices in the FSSM. 

\vspace*{-2.5mm}
\section{The Higgs sector of the FSSM}
\label{sec:spectrum}
\vspace*{-2.5mm}

In this section we describe the Higgs and higgsino sectors of the
FSSM, focusing on the hierarchical scenario in which the lightest
scalar plays the role of the SM Higgs boson, while the remaining
physical Higgs states are heavier.

The FSSM includes two $SU(2)$ doublets of chiral superfields with
positive hypercharge, $\hat{H}_u$ and $\hat{H}_u^\prime$, and two
doublets with negative hypercharge, $\hat{H}_d$ and
$\hat{H}_d^\prime$. The superpotential can be decomposed as $W = W_\mu
\,+ W_Y$, where $W_\mu$ generalizes the ``$\mu$ term'' of the MSSM:
\beq
W_\mu ~=~
\mu_{ud}\, \hat{H}_u \hat{H}_d ~+~
\mu_{u^\prime \!d^\prime}\, \hat{H}_{u}^\prime \hat{H}_d^\prime ~+~
\mu_{u^\prime \!d}\, \hat{H}_{u}^\prime \hat{H}_d ~+~
\mu_{u d^\prime}\, \hat{H}_{u} \hat{H}_d^\prime ~,
\label{eq:Wmu}
\eeq
whereas $W_Y$ contains the interactions of the Higgs doublets with the
quark and lepton superfields:
\beq
W_Y ~=~
- (Y_u \hat{H}_u + Y_u^\prime \hat{H}_{u}^\prime)\,\hat Q\hat{U}^c ~+~
(Y_d \hat{H}_d + Y_d^\prime \hat{H}_{d}^\prime)\,\hat Q\hat{D}^c ~+~
(Y_\ell \hat{H}_d + Y_\ell^\prime \hat{H}_{d}^\prime)\,\hat L\hat{E}^c~,
\eeq
where all gauge and generation indices are understood. In
ref.~\cite{Altmannshofer:2021hfu}, where the focus is on the leptonic
sector, the coupling $Y_\ell$ is defined as a rank-1 matrix whose only
non-zero element is $(3,3)$, providing a tree-level mass to the tau
lepton proportional to $v_d$. The coupling $Y_\ell^\prime$ is instead
defined as a rank-3 matrix which provides mass and mixing terms
proportional to $v_d^\prime$ to all of the charged leptons. In this
setup the muon Yukawa coupling can in principle be larger than the
bottom and tau ones, as long as $v_d^\prime \ll v_d$. As discussed in
ref.~\cite{Altmannshofer:2021hfu}, the current bounds on lepton-flavor
violating processes give rise to constraints on the off-diagonal
elements of $Y_\ell^\prime$, which anyway are not relevant to the
prediction for $a_\mu$ at the considered level of accuracy. Finally,
ref.~\cite{Altmannshofer:2021hfu} mentions that a similar construction
can be implemented in the quark sector.

In this work we do not consider flavor-violating processes in either
the lepton or the quark sector, but we rather focus on the interplay
of the effects of ${\cal O}(1)$ flavor-diagonal couplings on the
predictions for the SM-like Higgs mass and for $a_\mu$. We therefore
adopt for simplicity a pared-down version of $W_Y$, in which we
include only flavor-diagonal couplings for the second and third
generations:
\bea
W_Y &=&
-y_c^\prime \hat{H}_{u}^\prime \hat Q_2\hat{U}_2^c
~+~y_s^\prime \hat{H}_{d}^\prime \hat Q_2\hat{D}_2^c
~+~ y_\mu^\prime \hat{H}_{d}^\prime \hat L_2\hat{E}_2^c\nn\\[1mm]
&&-y_t^\prime \hat{H}_u^\prime \hat Q_3\hat{U}_3^c
~+~y_b^\prime \hat{H}_d^\prime \hat Q_3\hat{D}_3^c
~+~y_\tau^\prime \hat{H}_d^\prime \hat L_3\hat{E}_3^c\nn\\[1mm]
&&-y_t \hat{H}_u \hat Q_3\hat{U}_3^c
~+~y_b \hat{H}_d \hat Q_3\hat{D}_3^c
~+~y_\tau \hat{H}_d \hat L_3\hat{E}_3^c~.
\label{eq:WY}
\eea
As to the first-generation couplings, they are necessarily suppressed
with respect to those of the second generation, because in the FSSM
both generations receive their masses from $v_u^\prime$ and
$v_d^\prime$.

In addition to mass terms for gauginos and sfermions, which are the
same as in the MSSM, the soft SUSY-breaking Lagrangian of the FSSM
contains mass terms and $B$-terms for all of the Higgs doublets
\bea
-{\cal L}_{\rm {\scriptscriptstyle soft}} &\supset&
m^2_{uu}\,{H}_u^\dagger {H}_u ~+~ 
m^2_{dd}\,{H}_d^\dagger {H}_d ~+~
m^2_{u^\prime \!u^\prime}{H}_u^{\prime\dagger} {H}_u^\prime ~+~
m^2_{d^\prime \!d^\prime}{H}_d^{\prime\dagger} {H}_d^\prime \nn\\[2mm]
&+& \left(m^2_{uu^\prime}\,{H}_u^\dagger {H}_u^\prime
~+~ m^2_{dd^\prime}\,{H}_d^\dagger {H}_d^\prime ~+~ {\rm h.c.}\right)\nn\\[1mm]
&+& \left(B_{ud}\, {H}_u {H}_d ~+~
B_{u^\prime \!d^\prime}\, {H}_{u}^\prime {H}_d^\prime ~+~
B_{u^\prime \!d}\, {H}_{u}^\prime {H}_d^{\phantom \dagger} \,+~
B_{u d^\prime}\, {H}_{u} {H}_d^\prime ~+ {\rm h.c.}\right) ~,
\label{eq:LsoftH}
\eea
as well as trilinear interaction terms analogous to those in the superpotential
\bea
-{\cal L}_{\rm {\scriptscriptstyle soft}} &\supset&
-y_c^\prime A_c^\prime \,{H}_{u}^\prime  Q_2{U}_2^c
~+~y_s^\prime A_s^\prime \,{H}_{d}^\prime  Q_2{D}_2^c
~+~ y_\mu^\prime A_\mu^\prime \,{H}_{d}^\prime  L_2{E}_2^c\nn\\[1mm]
&&-y_t^\prime A_t^\prime \,{H}_u^\prime  Q_3{U}_3^c
~+~y_b^\prime A_b^\prime \,{H}_d^\prime  Q_3{D}_3^c
~+~y_\tau^\prime A_\tau^\prime \,{H}_d^\prime  L_3{E}_3^c\nn\\[1mm]
&&-y_t A_t \,{H}_u  Q_3{U}_3^c
~+~y_b A_b \,{H}_d  Q_3{D}_3^c
~+~y_\tau A_\tau \,{H}_d  L_3{E}_3^c~.
\label{eq:LsoftA}
\eea

The tree-level Higgs mass spectrum of a model with three pairs of
doublets has been discussed in ref.~\cite{Escudero:2005hk}, whose
approach can be easily adapted to the case of two pairs of doublets. In
order to identify the state that plays the role of the SM-like Higgs
boson, we rotate the four doublets to the so-called ``Higgs basis'',
in which only one of the doublets acquires a non-zero vev defined by
$v^2 \equiv v_u^2 + v_u^{\prime\,2} + v_d^2 + v_d^{\prime\,2}\,$. To
this purpose, we first rotate the doublets with the same hypercharge:
\beq
\left(\!\begin{array}{c} \Phi_u\\\Phi^\prime_u \end{array}\!\right) = 
\left(\!\begin{array}{rr} \sin \beta_u&\!\cos\beta_u\\\cos\beta_u
&\!-\sin\beta_u \end{array}\!\right)
\left(\!\begin{array}{c} H_u\\H^\prime_u \end{array}\!\right)~,~~~
\left(\!\begin{array}{c} \Phi_d\\\Phi^\prime_d \end{array}\!\right) = 
\left(\!\begin{array}{rr} \sin \beta_d&\!\cos\beta_d\\\cos\beta_d
&\!-\sin\beta_d \end{array}\!\right)
\left(\!\begin{array}{c} -\epsilon H^*_d\\-\epsilon H_d^{\prime\,*}
\end{array}\!\right)~,
\eeq
where the rotation angles are defined by $\tan\beta_u \equiv
v_u/v^\prime_u$ and $\tan\beta_d \equiv v_d/v^\prime_d$. The antisymmetric
tensor $\epsilon$, with $\epsilon_{12}=1$, acts on the complex
conjugates of $H_d$ and $H_d^\prime$ so that all doublets in the new
basis have the same hypercharge. In this basis, the vevs of the
neutral components of the four doublets become $\langle \Phi_u^0
\rangle = (v_u^2 + v_u^{\prime\,2})^{1/2}$, $\langle \Phi_d^0 \rangle
= (v_d^2 + v_d^{\prime\,2})^{1/2}$, and $\langle \Phi_u^{\prime \,0}
\rangle = \langle \Phi_d^{\prime \, 0} \rangle = 0$. The two doublets
that acquire vevs are further rotated as
\beq
\left(\!\begin{array}{c} \Phi_h \\ \Phi_\smallH \end{array}\!\right) =
\left(\!\begin{array}{rr} \cos \tilde
\beta&\!\sin\tilde\beta\\-\sin\tilde\beta
&\!\cos\tilde\beta \end{array}\!\right) \left(\!\begin{array}{c}
\Phi_d\\\Phi_u \end{array}\!\right)~,~~~~~ \tan\tilde\beta \equiv
\left(\frac{v_u^2 + v_u^{\prime\,2}}{v_d^2 +
  v_d^{\prime\,2}}\right)^{1/2},
\eeq
so that $\langle \Phi_h^0 \rangle = v$ and $\langle \Phi_\smallH^0
\rangle = 0$, i.e., in the Higgs basis the doublet $\Phi_h$ is
entirely responsible for the breaking of the EW symmetry (EWSB).

The mass matrices for the scalar, pseudoscalar and charged components
of the four doublets in the Higgs basis are given in the
appendix. They depend on the $\mu$ parameters defined in
eq.~(\ref{eq:Wmu}) and on the soft SUSY-breaking mass and $B$
parameters defined in eq.~(\ref{eq:LsoftH}), plus the EW gauge
couplings, the vev $v$ and the angles $\beta_u$, $\beta_d$ and
$\tilde\beta$. The minimum conditions of the scalar potential are used
to remove the dependence of the mass matrices on four combinations of
the original parameters. Most importantly, the terms that mix the
components of $\Phi_h$ with the components of the remaining doublets
are either zero or proportional to $v^2$ (more specifically, to
$\MZ^2$). In a hierarchical scenario in which the masses of the BSM
Higgs bosons are significantly higher than the EW scale, we can thus
neglect their mixing with $\Phi_h$, and identify the latter directly
with the Higgs boson of the SM.
In contrast, the scalar, pseudoscalar and charged components of the
three remaining doublets $\Phi_\smallH$, $\Phi_u^\prime$ and
$\Phi_d^\prime$ do mix with each other.\footnote{Note that in this
study we do not consider the possibility of CP violation in the Higgs
sector, hence the scalar and pseudoscalar components of the three
heavy doublets mix separately.}  However, under the approximation of
neglecting terms proportional to $v^2$, the respective $3\!\times\!3$
mass matrices are all the same. We can then combine the eigenstates of
the scalar, pseudoscalar and charged mass matrices into three heavy
doublets $H_i$ (with $i=1,2,3$), whose masses we denote as
$M_{H_i}$. The condition for their decoupling from the lightest
doublet is then $M_{H_i} \gg \MZ$.

\bigskip

We now focus on the properties of the SM-like doublet $\Phi_h$. The
tree-level mass of its scalar component is
\beq
(M_h^2)^{\rm {\scriptscriptstyle tree}} \,=~ 
\MZ^2\,\cos^2 2\tilde \beta~,
\eeq
which differs from the analogous result in the decoupling limit of the
MSSM only via the replacement of $\beta$ with $\tilde\beta$. In the
scenarios of interest for the solution to the $\gm2$ anomaly, one has
$v_d^\prime \ll v_d$. If the condition $v_u^\prime \ll v_u$ also holds,
$\tan\beta$ and $\tan\tilde\beta$ are numerically very close to each
other, hence the tree-level prediction for the SM-like Higgs mass in
the FSSM is essentially the same as in the MSSM.

The SM-like couplings of $\Phi_h$ to second-generation quarks and
leptons are related at the tree level to the superpotential couplings
in eq.~(\ref{eq:WY}) by
\beq
g_c ~=~ y_c^\prime \, \sin\tilde\beta \cos\beta_u~,~~~~~~
g_{s,\mu} ~=~ y_{s,\mu}^\prime \, \cos\tilde\beta \cos\beta_d~,
\label{eq:Yuk2}
\eeq
while the couplings to third-generation fermions read
\beq
g_t ~=~ y_t \, \sin\tilde\beta \sin\beta_u ~+~
y_t^\prime \, \sin\tilde\beta \cos\beta_u~,~~~~~~
g_{b,\tau} ~=~ y_{b,\tau} \, \cos\tilde\beta \sin\beta_d~+~
y_{b,\tau}^\prime \, \cos\tilde\beta \cos\beta_d~.
\label{eq:Yuk3}
\eeq
The relevant difference with the MSSM, in the context of the solution
of the $\gm2$ anomaly, is the additional suppression by $\cos\beta_d$
in the couplings of the SM-like Higgs to down-type fermions of the
second generation. Consequently, in the FSSM superpotential of
eq.~(\ref{eq:WY}), the muon Yukawa coupling can in principle be even
larger the bottom and tau ones, as long as $\tan\beta_d\gg1$.

For what concerns the couplings of $\Phi_h$ to sfermions, the quartic
couplings are proportional to the squared Yukawa couplings $g_f^2$
defined as in eqs.~(\ref{eq:Yuk2}) and (\ref{eq:Yuk3}). The main
difference with respect to the MSSM stems from the left-right mixing
parameters entering the trilinear Higgs-sfermion couplings in the
combination $g_f\,X_f$. Those for the second-generation sfermions read
\beq
X_c ~=~ A_c^\prime
\,-\, \cot\beta\,\tan\beta_u 
\left(\mu_{u^\prime\!d} + \mu_{u^\prime\!d^\prime}\cot\beta_d^{\phantom{\dagger}}
\right)~,~~~~~
X_{s,\mu} ~=~ A_{s,\mu}^\prime
\,-\, \tan\beta\,\tan\beta_d 
\left( \mu_{ud^\prime} + \mu_{u^\prime\!d^\prime}\cot\beta_u^{\phantom{\dagger}} \right)~,
\label{eq:Xf2}
\eeq
while those for the third-generation sfermions read

\bea
X_t & =& \frac{A_t
- \cot\beta \left( \mu_{ud}^{\phantom{\dagger}} +
\mu_{ud^\prime}\cot\beta_d\right)}
    {1 + \frac{y_t^\prime}{y_t}\cot\beta_u}~+~
\frac{A_t^\prime
-\cot\beta\,\tan\beta_u 
\left( \mu_{u^\prime \!d}
+ \mu_{u^\prime \!d^\prime}\cot\beta_d^{\phantom{\dagger}} \right)}
     {1 + \frac{y_t^{\phantom{\prime}}}{y_t^\prime}\tan\beta_u}~,\nn\\[2mm]
X_{b,\tau} &=& \frac{A_{b,\tau}
- \tan\beta 
\left( \mu_{ud}^{\phantom{\dagger}} +
\mu_{u^\prime \!d}\cot\beta_u\right)}
    {1 + \frac{y_{b,\tau}^\prime}{y_{b,\tau}}\cot\beta_d}~+~
\frac{A_{b,\tau}^\prime
- \tan\beta\,\tan\beta_d 
\left( \mu_{ud^\prime} +\mu_{u^\prime \!d^\prime}\cot\beta_u^{\phantom{\dagger}}
\right)}
     {1 + \frac{y_{b,\tau}^{\phantom{\prime}}}{y_{b,\tau}^\prime}\tan\beta_d}~.     
\label{eq:Xf3}
\eea
Again, the relevant aspect of eqs.~(\ref{eq:Xf2}) and (\ref{eq:Xf3})
in the context of the solution of the $\gm2$ anomaly is the
enhancement of the Higgs-smuon trilinear coupling by a factor
$\tan\beta_d$ with respect to the MSSM case (note that, to facilitate
the comparison, we expressed the trilinear couplings in terms of
$\tan\beta = \tan\tilde\beta \,\sin\beta_u/\sin\beta_d$). If we also
assume $\tan\beta_u\gg1$, the enhanced part of $X_\mu$ involves only
the superpotential parameter $\mu_{ud^\prime}$. We note, on the other
hand, that for $\tan\beta_{u,d}\gg1$ there are no further enhancements
with respect to the MSSM in the trilinear Higgs couplings to
third-generation sfermions, as long as the ``primed'' top, bottom and
tau Yukawa couplings remain at most of ${\cal O}(1)$.

\bigskip

We finally comment on the higgsino masses. In the hierarchical
scenario considered in our study, we assume that both gaugino and
higgsino masses are somewhat removed from the EW scale. In this case,
the mixing between EW gauginos and higgsinos induced by EWSB can be
neglected, and the four two-component fermions $\tilde h_u$, $\tilde
h_d$, $\tilde h_u^\prime$, and $\tilde h_d^\prime$ combine into two
Dirac fermions. Following ref.~\cite{Altmannshofer:2021hfu}, we define
the angles $\theta_u$ and $\theta_d$ that diagonalize the higgsino
mass matrix as
\beq
\left(\!\begin{array}{rr} \cos \theta_d&\!\sin\theta_d\\-\sin\theta_d
&\!\cos\theta_d \end{array}\!\right)
\left(\!\begin{array}{rr} \mu_{ud}&\! \mu_{u^\prime \!d} \\
\mu_{ud^\prime} &\! \mu_{u^\prime \!d^\prime} \end{array}\!\right)
\left(\!\begin{array}{rr} \cos \theta_u&\!\sin\theta_u\\-\sin\theta_u
&\!\cos\theta_u \end{array}\!\right)
~=~
\left(\!\begin{array}{rr} \mu &\!0\\0
&\!\tilde\mu \end{array}\!\right)~,
\eeq
and we use the Dirac masses $\mu$ and $\tilde \mu$ and the two
rotation angles as input parameters in our analysis. We note that the
numerical results in ref.~\cite{Altmannshofer:2021hfu} are obtained
for the parameter choices $\theta_u=\theta_d= \pi/4$ and $\tilde \mu =
\mu$, which in terms of the original superpotential parameters
correspond to $\mu_{ud^\prime} = - \mu_{u^\prime \!d} = \mu$ and
$\mu_{ud} = \mu_{u^\prime \!d^\prime} = 0$. While these choices might
look {\em ad hoc}, they are in fact quite appropriate, because they
make the dependence of the numerical results on $\mu_{ud^\prime}$ --
the parameter that determines the leading contributions from smuon
loops to both $a_\mu$ and the Higgs-mass correction -- more
transparent.

\section{Higgs-mass calculation in the EFT approach}
\label{sec:matching}

For our calculation of the radiative corrections to the Higgs mass in
the FSSM we adopt an EFT approach in which the effective theory valid
below the scale $\MS$ that characterizes the SUSY-particle masses is
just the SM. Rather than computing the prediction for the Higgs mass
from a full set of high-energy FSSM parameters, and then comparing it
with the value measured at the LHC, we follow a more convenient
procedure that uses the measured Higgs mass directly as an input
parameter. From the Higgs mass we extract the quartic Higgs coupling
$\lambdaSM$ at the EW scale, evolve it up to the SUSY scale with the
RGEs of the SM, and then require that $\lambdaSM(\MS)$ coincide with
the FSSM prediction for the quartic coupling of the lightest Higgs
scalar. This procedure allows us to determine one of the FSSM
parameters, such as, e.g., a common mass term for the stops.

We obtain a full one-loop prediction for the quartic coupling of the
SM-like Higgs doublet $\Phi_h$ in the FSSM. Combined with the one-loop
determination of the $\msbar$-renormalized parameters of the SM
Lagrangian at the EW scale, and with the two-loop RGEs of the
SM for the evolution up to the SUSY scale, this allows for the NLL
resummation of the corrections to the SM-like Higgs mass. However,
when they are available we use two-loop results for the determination
of the SM parameters and three-loop RGEs for their evolution. While in
the absence of a full two-loop calculation of the quartic coupling
this cannot be claimed to improve the overall accuracy of the
calculation, it does not degrade it either. Indeed, in the EFT
approach the EW-scale and SUSY-scale sides of the calculation are
separately free of large logarithmic corrections, and the inclusion of
additional pieces in only one side does not entail the risk of
spoiling crucial cancellations between large corrections.

\bigskip

We use the public code {\tt mr}~\cite{Kniehl:2016enc}, based on the
two-loop calculation of ref.~\cite{Kniehl:2015nwa}, to determine the
parameters of the SM Lagrangian -- neglecting all Yukawa couplings
except the top and bottom ones -- in the $\msbar$ renormalization
scheme at the scale $\qew = M_t$. We take as input for the code a set
of seven physical observables that we fix to their current PDG
values~\cite{ParticleDataGroup:2020ssz}, namely $G_F= 1.1663787 \times
10^{-5}$~GeV$^{-2}$, $M_h = 125.25$~GeV, $\MZ = 91.1876$~GeV, $\MW =
80.379$~GeV, $M_t= 172.76$~GeV, $M_b = 4.78$~GeV and
$\alpha_s(\MZ)=0.1179$. The remaining SM parameters that we need to
determine are the tau Yukawa coupling and the Yukawa couplings of the
second generation. For the leptons we take as input the physical
masses $M_\tau = 1.776$~GeV and $M_\mu = 105.66$~MeV, and obtain the
$\msbar$ Yukawa couplings directly at the scale $\qew = M_t$ via the
one-loop relation~\cite{Hempfling:1994ar}
\beq
g_\ell(\qew) ~=~ \sqrt{2 \sqrt{2}\, G_F}\, M_\ell\,\biggr[1 +
  \frac{\alpha}{4\pi}\left(3\, \ln\frac{M_\ell^2}{\qew^2}-4\right)
  +  \delta^{\rm{\scriptscriptstyle EW}}(\qew) \biggr]~,~~~~~(\ell = \tau,\, \mu)~,
\eeq
where the EW correction stemming from the renormalization of $G_F$
reads\,\footnote{We use here an approximate formula from
ref.~\cite{Hempfling:1994ar} which includes only the contributions
from the top Yukawa coupling and the quartic Higgs coupling. Anyway,
the overall effect of this correction is only about $0.5\%$.}
\beq
\delta^{\rm{\scriptscriptstyle EW}}(\qew) ~=~
\frac{G_F}{8\pi^2\sqrt2}
\left[3\,M_t^2\,\left(\frac12 - \ln\frac{M_t^2}{\qew^2}\right)
  \,+\,\frac{M_h^2}{4}\,\right]~.
\label{eq:dEW}
\eeq
For the second-generation quarks we take as input the
$\msbar$-renormalized masses $m_c(m_c)=1.27$~GeV and $m_s(2~{\rm
  GeV})=93$~MeV~\cite{ParticleDataGroup:2020ssz}, which we evolve up
to the scale $\qew$ at the NLL level in QCD by means of eqs.~(D4) and
(D5) of ref.~\cite{Degrassi:2007kj}. We then include the one-loop QED
and EW corrections according to:
\beq
g_q(\qew) ~=~ \sqrt{2 \sqrt{2}\, G_F}\, m_q(\qew)\,\biggr[1 +
 \frac{3 \,\alpha}{4\pi}\,Q_q^2\, \ln\frac{m_q^2}{\qew^2}
  +  \delta^{\rm{\scriptscriptstyle EW}}(\qew) \biggr]~,~~~~~(q = c,\,s)~,
\eeq
where $Q_q$ is the electric charge of the quark $q$, and
$\delta^{\rm{\scriptscriptstyle EW}}(\qew)$ is given in
eq.~(\ref{eq:dEW}).

\bigskip

For the evolution of the SM couplings from the EW scale to the SUSY
scale we use the set of three-loop RGEs provided in
refs.~\cite{Bednyakov:2012rb, Bednyakov:2012en, Bednyakov:2013eba},
which however include only the third-generation Yukawa couplings. For
the couplings of the second generation we use 2-loop RGEs from
ref.~\cite{Luo:2002ey}, which are sufficient to our aim of a NLL
resummation of the large logarithmic effects. Following
refs.~\cite{Bednyakov:2012rb, Bednyakov:2012en, Bednyakov:2013eba}, we
neglect the tiny contributions of the Yukawa couplings of the first
two generations within the beta functions, apart from the overall
multiplicative factors.

\bigskip

Once the SM couplings are evolved up to the SUSY scale, they are
matched to the corresponding FSSM couplings, which enter the
prediction for the quartic Higgs coupling. Since the Yukawa couplings
enter only from one loop onwards, the tree-level relations in
eqs.~(\ref{eq:Yuk2}) and (\ref{eq:Yuk3}) are in principle sufficient
for the NLL calculation of the Higgs-mass prediction. It is
nevertheless convenient to take into account the one-loop ``SUSY-QCD''
corrections controlled by the strong gauge coupling $g_3$ to the
relation between the quark Yukawa couplings of the SM and those of the
FSSM. This amounts to redefining the quark Yukawa couplings as
\beq
\label{eq:gqhat}
\hat g_q(Q) ~=~ \frac{g_q(Q)}{1-\Delta g_q}~,~~~~~(q=t,\,b,\,c,\,s)~,
\eeq
where $g_q(Q)$ are given in eqs.~(\ref{eq:Yuk2}) and (\ref{eq:Yuk3}),
and the correction $\Delta g_q$ reads
\beq
\label{eq:deltag}
\Delta g_q  ~=~ - \frac{g_3^2}{12\pi^2}\,\left[
\,1+\,\ln\frac{M_3^2}{Q^2} 
\,+\, \wt F_6\left(\frac{M_{\tilde q_\smallL}}{M_3}\right)
\,+\, \wt F_6\left(\frac{M_{\tilde q_\smallR}}{M_3}\right)
\,-\, \frac{X_q}{M_3}\,
\wt F_9\left(\frac{M_{\tilde q_\smallL}}{M_3},
\frac{M_{\tilde q_\smallR}}{M_3}\right)\right]~,
\eeq
where: $M_3$ is the gluino mass; $M_{\tilde q_\smallL}$ and $M_{\tilde
  q_\smallR}$ are the soft SUSY-breaking mass parameters for the
scalar partners of the left- and right-handed quarks, respectively;
$X_q$ are the left-right mixing parameters given in
eqs.~(\ref{eq:Xf2}) and (\ref{eq:Xf3}); the functions $\wt F_6(x)$ and
$\wt F_9(x,y)$ are defined in the appendix A of
ref.~\cite{Bagnaschi:2014rsa}. As was recently discussed in a
systematic way in refs.~\cite{Kwasnitza:2020wli, Kwasnitza:2021idg}
for the case of the MSSM, the use of the corrected Yukawa couplings
$\hat g_q$ absorbs (``resums'') in the one-loop contribution to the
quartic Higgs coupling a tower of higher-order corrections involving
powers of $g_3^2\,X_q/\MS$, where $\MS$ denotes the scale of the
squark and gluino masses. In case $X_q/\MS$ contains terms that are
numerically enhanced (e.g., by a large ratio of vevs), such
``resummation'' ensures a better convergence of the perturbative
expansion. We note that contributions to $\Delta g_q$ controlled by
the Yukawa couplings and by the EW gauge couplings also exist, but we
do not consider them in our study as they are generally subdominant to
those controlled by the strong gauge coupling.\footnote{This is not
necessarily the case for the $\tan\beta$-enhanced ${\cal O}(g_t^2)$
contribution to $\Delta g_b$, but in the FSSM that contribution
depends on $\mu_{ud}$, and vanishes in the scenarios considered in
this paper.}

For what concerns the Yukawa couplings of the leptons, the only
one-loop corrections that can be enhanced by a large ratio of vevs are
those controlled by the EW gauge couplings. In particular, in the FSSM
the muon Yukawa coupling is subject to corrections enhanced by
$v_u/v_d^\prime = \tan\beta\tan\beta_d$, which, following
ref.~\cite{Altmannshofer:2021hfu}, we absorb in the coupling via the
redefinition
\beq
\label{eq:gmuhat}
\hat g_\mu(Q) ~=~ \frac{g_\mu(Q)}{1+\epsilon_\ell\,\tan\beta\tan\beta_d}~,
\eeq
where the explicit formula for $\epsilon_\ell$ is given in
ref.~\cite{Altmannshofer:2021hfu}. Being controlled by the EW gauge
couplings, the term $\epsilon_\ell$ is itself of ${\cal O}(10^{-3})$
only, but the overall correction in eq.~(\ref{eq:gmuhat}) is not
negligible for the values of $\tan\beta\tan\beta_d$ in the few-hundred
range that -- as will be seen in section \ref{sec:numerical} -- are
relevant to the solution of the $\gm2$ anomaly. For the tau Yukawa
coupling, on the other hand, the analogous EW correction is enhanced
at most by $\tan\beta$, still with an ${\cal O}(10^{-3})$
prefactor. We can thus neglect this correction in our analysis and
define $\hat g_\tau(Q) = g_\tau(Q)$.

At this stage, an issue with the SUSY-QCD correction to the strange
Yukawa coupling might be worth mentioning. Inspection of
eq.~(\ref{eq:Xf2}) shows that the trilinear Higgs-squark coupling
$X_s$ entering the correction $\Delta g_s$ in eq.~(\ref{eq:deltag})
includes the term $\mu_{ud^\prime}\tan\beta\tan\beta_d$\,, which is
the same combination of parameters entering the dominant
higgsino-gaugino-smuon contribution to $a_\mu$. Being controlled by
the strong gauge coupling, $\Delta g_s$ is of the order of
$10^{-2}\times \tan\beta\tan\beta_d$\,, and can easily reach and even
exceed unity for the values of $\tan\beta\tan\beta_d$ relevant to the
solution of the $\gm2$ anomaly. A particularly obnoxious situation
occurs when $\Delta g_s \simeq 1$, in which case the corrected
coupling $\hat g_s$ in eq.~(\ref{eq:gqhat}) blows up, leading to
unphysically large corrections and numerical instabilities. Since the
higgsino-gaugino-smuon contribution to $a_\mu$ takes the sign required
to account for the observed anomaly when $(\mu_{ud^\prime}M_2 )>0$,
the condition $\Delta g_s \simeq 1$ requires $(M_2 M_3)<0$, as is the
case in scenarios with anomaly mediated SUSY breaking
(AMSB).\footnote{In the case of the MSSM with AMSB, the interplay
between contributions to $\gm2$ and SUSY-QCD corrections to the quark
couplings was discussed earlier in ref.~\cite{Allanach:2009ne}.} Even
in scenarios where the SUSY-QCD correction suppresses $\hat g_s$
rather than enhancing it, the condition $|\Delta g_s|>1$ means that
the radiative correction to the strange quark mass arising from
squark-gluino diagrams exceeds, and possibly by far, the tree-level
contribution. This would complicate any attempt (which we do not make
in this paper anyway) to obtain a realistic flavor structure for the
quark sector of the FSSM. We remark that a trivial way out from this
complication would consist in applying the four-doublet construction
only to the lepton sector, and have all of the quarks receive their
masses from the doublets $H_u$ and $H_d$.

\bigskip

We now describe the one-loop matching condition for the quartic Higgs
coupling in the FSSM. At a renormalization scale $Q$ of the order of
the SUSY particle masses, it takes the form
\beq
\label{eq:lambda1loop}
\lambdaSM(Q) ~=~
\frac14\,\left[g^2(Q) + \gpq(Q)\right] \,\cos^22\tilde\beta
~+~\Delta\lambda^{\rm reg} 
~+~\Delta\lambda^{\tilde f}
~+~\Delta\lambda^{\smallH}
~+~\Delta\lambda^{\chi}~,
\eeq
where $g$ and $\gp$ are the EW gauge couplings. Again, we see
that the tree-level matching condition differs from the analogous
result in the MSSM only via the replacement of $\beta$ with
$\tilde\beta$.  We assume that the EW gauge couplings are SM
parameters renormalized in the $\msbar$ scheme, i.e.~we use directly
the values obtained via RG evolution from the EW scale. Following
ref.~\cite{Bagnaschi:2014rsa}, we also assume that the angle
$\tilde\beta$ is renormalized in such a way as to remove entirely the
wave-function-renormalization (WFR) contributions that mix the SM-like
Higgs doublet with the heavy doublets.

The one-loop correction $\Delta\lambda^{\rm reg}$ accounts for the
fact that SUSY determines the quartic Higgs coupling in the $\drbar$
scheme, whereas $\lambdaSM$ and the EW gauge couplings in
eq.~(\ref{eq:lambda1loop}) are defined in the $\msbar$ scheme. It
reads~\cite{Bagnaschi:2014rsa}
\beq
\label{eq:dreg}
(4\pi)^2 \Delta\lambda^{\rm reg}~=~ -\frac{~\gpqq} 4\,
\,-\,\frac{~g^2 \gpq}2 \,-\,\left(\frac34
-\frac{\cos^22\tilde\beta}6\right)\,g^4~.
\eeq

Concerning the remaining one-loop threshold corrections in
eq.~(\ref{eq:lambda1loop}), $\Delta\lambda^{\tilde f}$ arises from
diagrams that involve the sfermions, $\Delta\lambda^{\smallH}$ from
diagrams that involve the heavy Higgs doublets, and
$\Delta\lambda^{\chi}$ from diagrams that involve higgsinos and EW
gauginos. Each of these three corrections can in turn be decomposed as
a sum of three terms:
\beq
\label{eq:dlambda}
\Delta\lambda^p ~=~
\Delta\lambda^{p,\,{\rm{\scriptscriptstyle 1PI}}} \,+\, 
\Delta\lambda^{p,\,{\rm{\scriptscriptstyle WFR}}} \,+\, 
\Delta\lambda^{p,\,{\rm{\scriptscriptstyle gauge}}}~,
~~~~~~(p = {\tilde f\,,H\,,\chi})~.
\eeq
The first term on the r.h.s.~of the equation above denotes the
contribution of one-particle-irreducible (1PI) diagrams with particles
of type $p$ in the loop and four external Higgs fields; the second
term involves the contributions of particles of type $p$ to the WFR of
the Higgs field, which multiply the tree-level quartic coupling; the
third term contains additional corrections stemming from the fact that
the SUSY prediction for the quartic Higgs coupling involves the gauge
couplings of the MSSM, whereas we interpret the gauge couplings in the
tree-level part of eq.~(\ref{eq:lambda1loop}) as SM parameters.

To obtain the four-Higgs diagrams entering
$\Delta\lambda^{p,\,{\rm{\scriptscriptstyle 1PI}}}$ and the
self-energy diagrams entering
$\Delta\lambda^{p,\,{\rm{\scriptscriptstyle WFR}}}$ we use the general
results from ref.~\cite{Braathen:2018htl} (see sections B.3 and B.1.1,
respectively, of that paper). This saves us the trouble of actually
calculating one-loop Feynman diagrams, but requires that we adapt to
the case of the FSSM the notation of ref.~\cite{Braathen:2018htl} for
masses and interactions of scalars and fermions in a general
renormalizable theory. The additional corrections in
$\Delta\lambda^{p,\,{\rm{\scriptscriptstyle gauge}}}$ can instead be
obtained by adapting the MSSM shifts of the gauge couplings, see
eqs.~(19) and (20) of ref.~\cite{Bagnaschi:2014rsa}, to the FSSM case
of two Dirac higgsinos with masses $\mu$ and $\tilde \mu$, and three
heavy Higgs doublets with masses $M_{H_i}$.

We find that the sfermion contribution to the quartic Higgs coupling,
$\Delta\lambda^{\tilde f}$, has the same form as the corresponding
contribution in the MSSM, see eq.~(A1) of
ref.~\cite{Bagnaschi:2017xid}, trivially extended to the case of
non-zero Yukawa couplings for the second generation. However, in the
FSSM case the angle $\beta$ is replaced by $\tilde\beta$, and the
trilinear Higgs-sfermion couplings $X_f$ are those given in our
eqs.~(\ref{eq:Xf2}) and (\ref{eq:Xf3}). In contrast, the heavy-Higgs
and higgsino-gaugino contributions differ from the corresponding MSSM
contributions, due to the extended Higgs/higgsino sector of the
FSSM. The full formulas for $\Delta\lambda^{\smallH}$ and, especially,
$\Delta\lambda^{\chi}$ for generic values of all relevant parameters
are lengthy and not particularly illuminating, therefore we make them
available on request in electronic form. In the following we provide
instead explicit results for all three contributions in the simplified
FSSM scenario that we will use in section~\ref{sec:numerical} to
explore the interplay between the prediction for the Higgs mass and
the solution of the $\gm2$ anomaly.

In the sfermion sector, we assume degenerate soft SUSY-breaking mass
parameters $M_{\tilde f_{12}}$ for all first- and second-generation
sfermions and $M_{\tilde f_3}$ for all third-generation sfermions. The
sfermion contribution to the quartic Higgs coupling then becomes:
\bea
(4\pi)^2\,\Delta\lambda^{\tilde f} &=&
\left[2\left(3\,\hat g_c^4+3\,\hat g_s^4+ \hat g_\mu^4\right)
+\frac{\,\bar g^2}2\left(3\,\hat g_c^2-3\,\hat g_s^2- \hat g_\mu^2\right)
\cos2\tilde\beta
+\frac23\left(g^4 + \frac53\,\gpqq\right)\cos^2 2\tilde\beta\right]
\ln\frac{M_{\tilde f_{12}}^2}{Q^2}\nn\\[2mm]
&+&
\left[2\left(3\,\hat g_t^4+3\,\hat g_b^4+ \hat g_\tau^4\right)
+\frac{\,\bar g^2}2\left(3\,\hat g_t^2-3\,\hat g_b^2- \hat g_\tau^2\right)
\cos2\tilde\beta
+\frac13\left(g^4 + \frac53\,\gpqq\right)\cos^2 2\tilde\beta\right]
\ln\frac{M_{\tilde f_{3}}^2}{Q^2}\nn\\[2mm]
&+& \sum_{f=c,s,\mu}\,\hat g_f^2\,N_c\,\frac{X_f^2}{M_{\tilde f_{12}}^2}
\left[2\,\hat g_f^2\left(1-\frac{X_f^2}{12\,M_{\tilde f_{12}}^2}\right)
  \,+\, \frac{\bar g^2}{12}\,\cos2\tilde\beta\,
  \left(3\,c_f-\cos2\tilde\beta\right)\right]\nn\\[2mm]
&+& \sum_{f=t,b,\tau}\,\hat g_f^2\,N_c\,\,\,\frac{X_f^2}{M_{\tilde f_{3}}^2}\,
\left[2\,\hat g_f^2\left(1-\,\frac{X_f^2}{12\,M_{\tilde f_{3}}^2}\,\right)
  \,+\, \frac{\bar g^2}{12}\,\cos2\tilde\beta\,
  \left(3\,c_f-\cos2\tilde\beta\right)\right]~,
\label{eq:dlsfe}
\eea
where $\hat g_f$ are the loop-corrected Yukawa couplings defined in
eqs.~(\ref{eq:gqhat})--(\ref{eq:gmuhat}), the trilinear Higgs-sfermion
couplings $X_f$ are given in eqs.~(\ref{eq:Xf2}) and (\ref{eq:Xf3}),
and we defined: $\bar g^2 \equiv g^2 + \gpq$\,; $N_c=3$ for quarks and
$N_c=1$ for leptons; $c_f =1 $ for $f=c,t$ and $c_f=-1$ for
$f=s,b,\mu,\tau$.

\bigskip

In the Higgs sector, we assume that there is no mixing between the
three doublets $\Phi_H$, $\Phi_u^\prime$ and $\Phi_d^\prime$, i.e.~we
take the $3\!\times\!3$ matrix $R_H$ that rotates the heavy doublets
from the Higgs basis to the basis of mass eigenstates to be the
identity. We also assume a common mass $M_H$ for all three of the
doublets. The heavy-Higgs contribution to the quartic Higgs coupling
then becomes:
\beq
(4\pi)^2\,\Delta\lambda^{\smallH} ~=~
\frac1{64} \left[16\,g^4 + 8\,\gpqq+ 7\,\bar g^4
  -4\,\left(\bar g^4-2\,\gpqq\right)\cos4\tilde \beta
  - 3\,\bar g^4 \cos8\tilde \beta\right]
\ln\frac{M_{H}^2}{Q^2}
~-~\frac{3\,\bar g^4}{16}\,\sin^2 4\tilde\beta~.
\label{eq:dlhig}
\eeq

\bigskip

Finally, for the higgsino sector we consider the same scenario as in
ref.~\cite{Altmannshofer:2021hfu}, namely $\theta_u=\theta_d= \pi/4$
and $\tilde \mu = \mu\,$, so that our choices for $\mu$ determine
directly the relevant parameter $\mu_{ud^\prime}$. We also assume a
common mass $M_\chi$ for the higgsinos and the EW gauginos,
i.e.~$M_\chi\equiv M_1=M_2=\mu=\tilde\mu$. The higgsino-gaugino
contribution to the quartic Higgs coupling then becomes:
\bea
(4\pi)^2\,\Delta\lambda^{\chi} &=&
-\frac{1}{24}\,\left[47\,g^4+12\,g^2\gpq+ 13\, \gpqq
  + \left(11\, g^4 - 12\, g^2\gpq + \gpqq\right) \cos 4\tilde \beta\right]
\ln\frac{M_{\chi}^2}{Q^2}\nn\\[2mm]
&-& \frac1{48}\,\biggr[
\,93\,g^4+50\,g^2\gpq+27\,\gpqq
+\left(
  3\,g^4 - 10\, g^2\gpq - 3\,\gpqq\right)\, \cos 4\tilde\beta
\nn\\[1mm]
&&~~~~~~~
 + 2\,\left(3\,g^4+4\,g^2\gpq+\gpqq\right)
 \,\cos 4\tilde\beta\,\sin2\tilde\beta\,\sin(\beta_d-\beta_u)\nn\\[2mm]
&&~~~~~~~ 
  -2\,\left(45\,g^4+28\, g^2\gpq + 15\,\gpqq\right)
  \,\sin2\tilde\beta\,\sin(\beta_d-\beta_u)\nn\\[1mm]
&&~~~~~~~
-2\,\left(3\,g^4+2\, g^2\gpq + \gpqq\right)
  \sin^22\tilde\beta\,\cos 2(\beta_d-\beta_u)  \biggr]~.
\label{eq:dlchi}
\eea

\bigskip

The inspection of eqs.~(\ref{eq:dlhig}) and (\ref{eq:dlchi}) shows
that the heavy-Higgs and higgsino-gaugino contributions to the quartic
Higgs coupling all involve four powers of the EW gauge
couplings. Their numerical impact is thus going to be modest, unless
there is a significant hierarchy between the matching scale $Q$ and
the mass scales $M_H$ and $M_\chi$. In contrast, the sfermion
contributions include terms depending on the top Yukawa coupling $\hat
g_t$, which is of ${\cal O}(1)$, as well as terms involving other
Yukawa couplings in which the smallness of $\hat g_f$ can be
compensated by a large ratio $X_{\tilde f}/M_{\tilde f}$. In
particular, the contribution that will be relevant to our discussion
in section~\ref{sec:numerical} is the one involving the muon Yukawa
coupling, which for $\tan\tilde\beta\gg1$ reads
\beq
\label{eq:dlsmu}
(4\pi)^2\,\Delta\lambda^{\tilde \mu} ~\approx~
\hat g_\mu^2\,\frac{X_\mu^2}{M_{\tilde \mu}^2}
\left[2\,\hat g_\mu^2\left(1-\frac{X_\mu^2}{12\,M_{\tilde \mu}^2}\right)
  \,+\, \frac{\bar g^2}{6} \right]~,
\eeq
where $\hat g_\mu$ and $X_\mu$ are defined in eqs.~(\ref{eq:gmuhat})
and (\ref{eq:Xf2}), respectively, and by $M_{\tilde \mu}$ we denote a
common mass parameter for the scalar partners of the left- and
right-handed muons (note that $M_{\tilde \mu} = M_{\tilde f_{12}}$ in
our simplified scenario). We recall that $X_\mu$ contains a term
enhanced by $\tan\beta\tan\beta_d$, and indeed when the combination
$(\mu_{ud^\prime}/M_{\tilde \mu})\tan\beta\tan\beta_d$ is large enough
to overcome the smallness of $\hat g_\mu$ the smuon contribution to
the quartic Higgs coupling becomes large and negative. As will be
discussed in the next section, an increased positive contribution from
a different SUSY sector is then necessary to maintain the correct
prediction for the SM-like Higgs mass. In particular, the stop
contribution to the quartic Higgs coupling is dominated by the terms
involving four powers of $\hat g_t$, which read
\beq
\label{eq:dlstop}
(4\pi)^2\,\Delta\lambda^{\tilde t} ~\approx~ 
6\, {\hat g}_t^4\,
\left(\ln\frac{M_{\tilde t}^2}{Q^2} + \frac{X_t^2}{M_{\tilde t}^2} -\frac{X_t^4}{12\,M_{\tilde t}^4}\right)~,
\eeq
where by $M_{\tilde t}$ we denote a common mass parameter for the
scalar partners of the left- and right-handed top (with $M_{\tilde t}
= M_{\tilde f_{3}}$ in our simplified scenario) and $X_t$ is defined
in eq.~(\ref{eq:Xf3}). The non-logarithmic terms in
eq.~(\ref{eq:dlstop}) are maximized for $X_t = \sqrt 6 \,M_{\tilde
  t}$\,, and a further increase in $\Delta\lambda^{\tilde t}$ can
arise from the logarithmic term when the stop mass is pushed to higher
values.

Finally, we note that in the FSSM the strange-squark contribution to
the quartic Higgs coupling is subject to the same enhancement by
$\tan\beta\tan\beta_d$ as the smuon contribution. However, the
strange-squark contribution is generally subdominant, because the
strange Yukawa coupling is smaller than the muon one at the matching
scale. A possible exception is the pathological case discussed
earlier, in which a SUSY correction $\Delta g_s \simeq 1$ in
eq.~(\ref{eq:gqhat}) causes the strange coupling $\hat g_s$ to blow
up.

\section{Higgs-mass constraints and $\gm2$}
\label{sec:numerical}

We now investigate the interplay between the constraints on the FSSM
parameter space arising from the solution to the $\gm2$ anomaly and
those arising from the prediction for the quartic Higgs coupling. To
keep the number of independent parameters manageable, we employ the
simplifying assumptions for the SUSY mass spectrum described in the
previous section. Namely, we adopt common mass scales $M_{\tilde
  f_{12}}$, $M_{\tilde f_{3}}$, $M_H$ and $M_\chi$ for
first/second-generation sfermions, third-generation sfermions, heavy
Higgs bosons and higgsinos/EW-gauginos, respectively. Note that we
will henceforth refer to the common mass parameters for the sfermions
as $M_{\tilde\mu}$ and $M_{\tilde t}$, because those are the masses of
the first/second and third generation, respectively, that are most
relevant to our discussion of $\gm2$ and of the Higgs mass
constraint. We will also keep referring to the collective scale of the
SUSY particle masses as $\MS$. A further simplifying assumption
consists in neglecting all contributions from the ``primed'' Yukawa
couplings for the third generation, namely $y_t^\prime$, $y_b^\prime$,
and $y_\tau^\prime$ in eq.~(\ref{eq:WY}), as they do not give rise to
contributions enhanced by large ratios of vevs. Finally, we take
directly as input the stop mixing parameter $X_t$, which enters the
stop contribution to $\Delta \lambda^{\tilde f}$, thereby fixing via
eq.~(\ref{eq:Xf3}) the soft SUSY-breaking trilinear coupling $A_t$ as
a function of the other parameters.\footnote{Our assumption
$y_t^\prime=y_b^\prime=y_\tau^\prime=0$ implies that the second term
on the r.h.s.~of each line of eq.~(\ref{eq:Xf3}) vanishes.} For the
remaining trilinear couplings in eq.~(\ref{eq:LsoftA}), which are not
involved in any enhanced contributions to $\Delta \lambda^{\tilde f}$,
we assume $A^\prime_c=A^\prime_s=A^\prime_\mu=0$ and $A_b=A_\tau=A_t$.

\bigskip

In the FSSM, $\gm2$ receives contributions from one-loop diagrams
involving smuons, higgsinos and EW gauginos that are enhanced by
$v_u/v_d^\prime = \tan\beta\tan\beta_d$. Explicit formulas for these
contributions with full dependence on the relevant FSSM parameters --
but under the assumption $\MS \gg v$, i.e.~neglecting the effects of
EWSB on the SUSY-particle masses and mixing -- are given
ref.~\cite{Altmannshofer:2021hfu}. In the simplified scenario
considered here, where in particular $M_{\tilde \mu_\smallL} =
M_{\tilde \mu_\smallR} = M_{\tilde \mu}$, $\mu=\tilde\mu =
M_1=M_2=M_\chi$, and $\theta_u=\theta_d=\pi/4$, they reduce to
\beq
\Delta a_\mu^\smallFSSM ~=~ \frac{1}{192\pi^2}\,
\frac{M_\mu^2}{M^2_{\tilde \mu}}\,
\frac{\tan\beta\tan\beta_d}{1+ \epsilon_\ell\,\tan\beta\tan\beta_d}\,
\biggr[\, \gpq \,f_1\left({M_\chi^2}/{M_{\tilde \mu}^2}\right)
  \,+\,5\,g^2 \,f_2\left({M_\chi^2}/{M_{\tilde \mu}^2}\right)\biggr]~,
\label{eq:damuFSSM}
\eeq
where
\bea
f_1(x) &=& ~~\frac{6x}{(1-x)^4}\,\biggr[
  7 + 4\,x - 11\,x^2 + 2\,(1+6\,x+ 2\,x^2)\,\ln x\biggr]~,~~~~~\,f_1(1)=1~,\\[2mm]
f_2(x) &=& \frac{6}{5\,(1-x)^4}\,\biggr[
  4 + 11\,x - 16\,x^2 + x^3 + 2\,x\,(7+2\,x)\,\ln x\biggr]~,~~~~f_2(1)=1~,
\label{eq:f2}
\eea
and $\epsilon_\ell$ represents the correction to the muon Yukawa
coupling introduced in eq.~(\ref{eq:gmuhat}). In our simplified
scenario, the formula given in ref.~\cite{Altmannshofer:2021hfu} for
this correction reduces to
\beq
\epsilon_\ell ~=~ \frac{\gpq}{64\pi^2}\,
g_1\left({M_\chi^2}/{M_{\tilde \mu}^2}\right)
\,-\,\frac{3\,g^2}{64\pi^2}\,
g_2\left({M_\chi^2}/{M_{\tilde \mu}^2}\right)~,
\label{eq:epsilon}
\eeq
where
\beq
g_1(x)\,=\,\frac{2x}{(1-x)^2}\biggr[3-3\,x+(1+2\,x)\,\ln x\biggr]~,~~~~~~
g_2(x)\,=\,\frac{2x}{(1-x)^2}\biggr[-1+x-\ln x\biggr]~,~~~~~~
g_1(1)=g_2(1)=1~.
\eeq

The requirement that the smuon-higgsino-gaugino contribution in
eq.~(\ref{eq:damuFSSM}) provide the solution of the $\gm2$ anomaly
corresponds to $\Delta a_\mu^\smallFSSM = 251\times 10^{-11}$. For a
given choice of values of $M_{\tilde \mu}$ and $M_\chi$, this can be
solved for the product $\tan\beta\tan\beta_d$, which in turn
determines the enhancement of the mixing parameter $X_\mu$ entering
the smuon contribution to $\Delta \lambda^{\tilde f}$. The requirement
that the FSSM prediction for the quartic Higgs coupling agree with the
value obtained by evolving $\lambdaSM$ from the EW scale up to the
SUSY scale can then be used to determine one of the remaining FSSM
parameters.  In particular, it seems reasonable to determine one of
the parameters that enter the dominant contribution to $\Delta
\lambda^{\tilde f}$, i.e.~the one involving the stops.

\begin{figure}[t]
\begin{center}
  \vspace*{-1.2cm}
  \includegraphics[width=8cm]{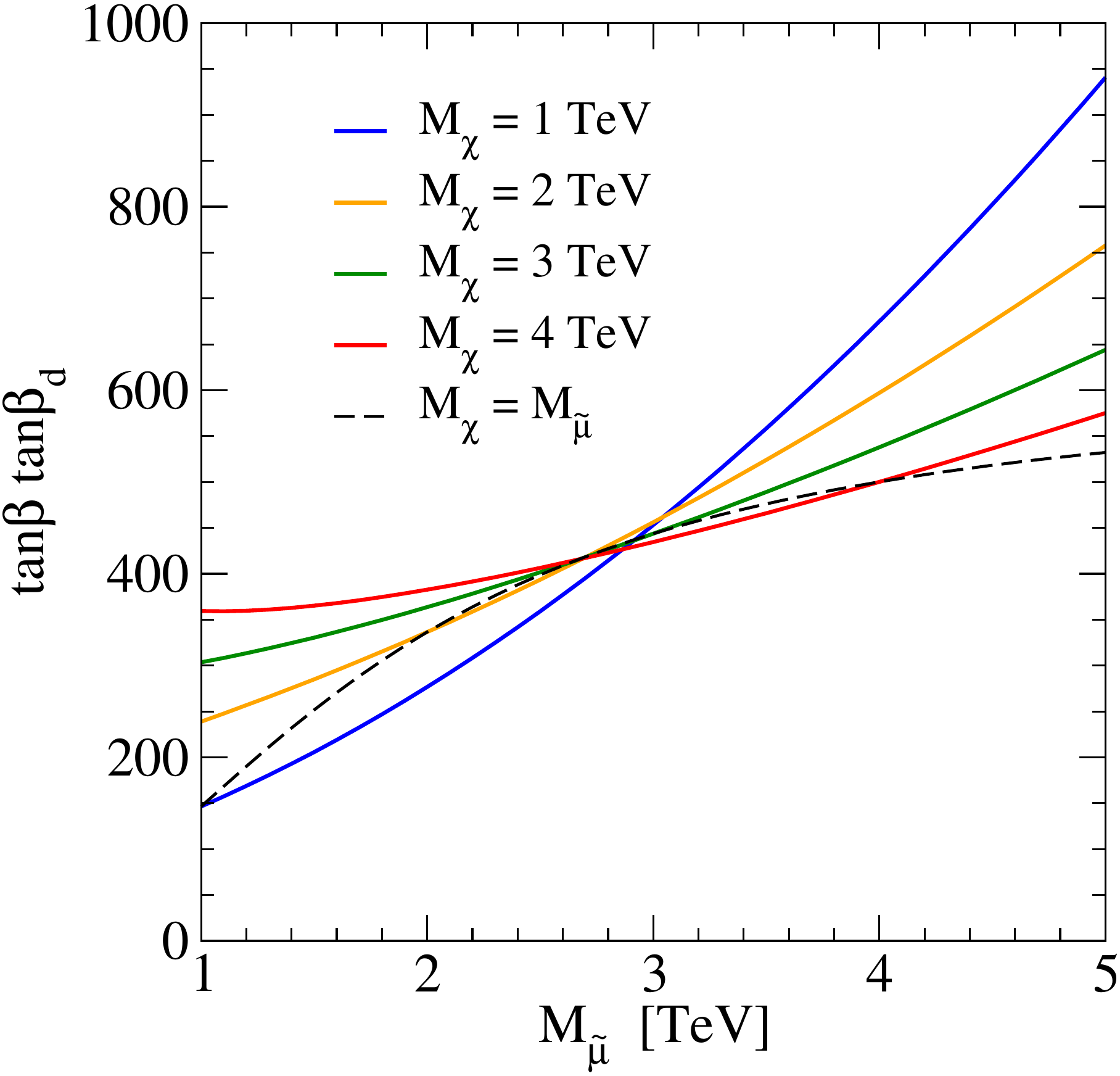}~~~~~
  \includegraphics[width=7.8cm]{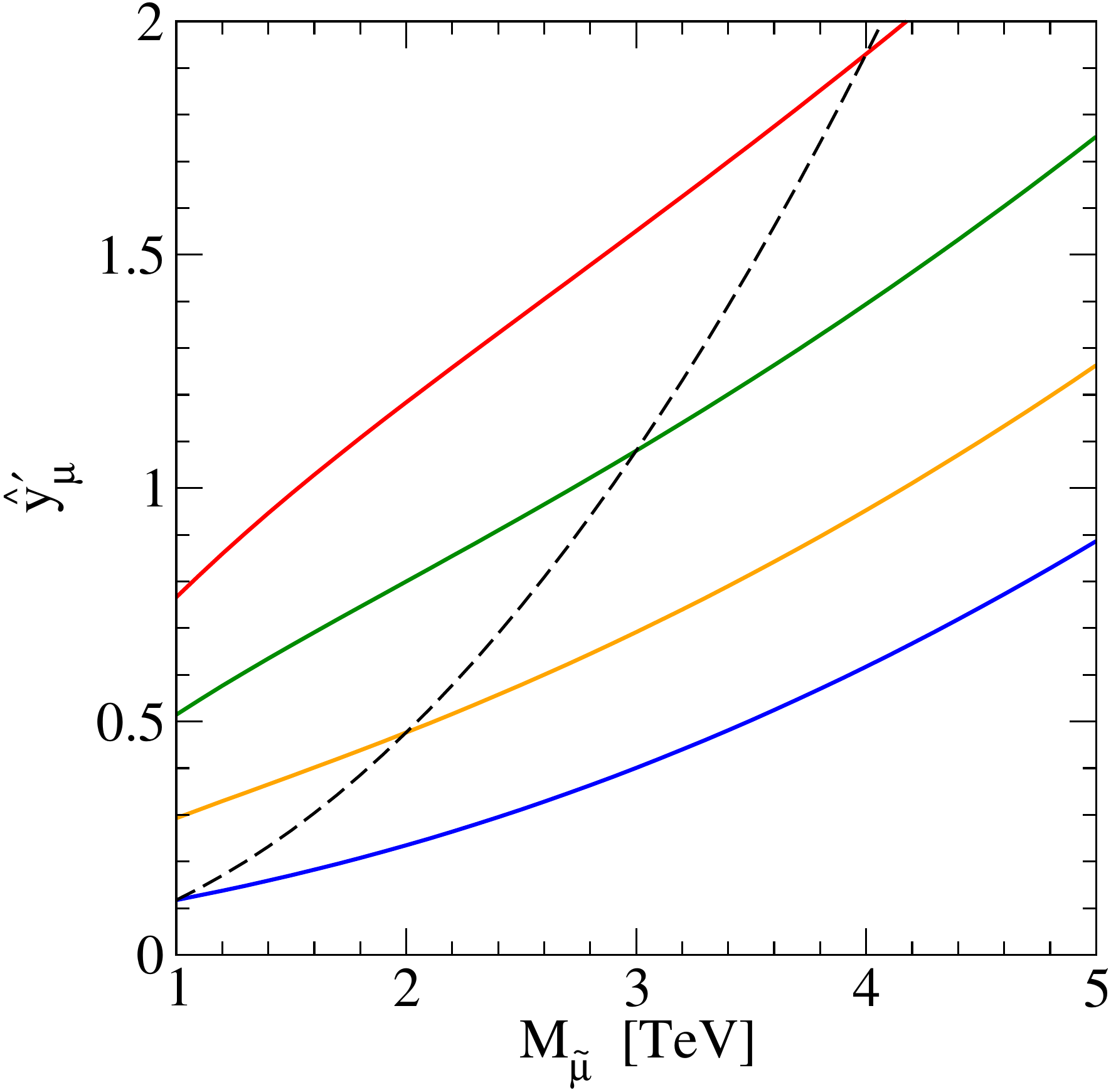}
  \caption{\em Left: Values of the product $\tan\beta\tan\beta_d$ that
    result in $\Delta a_\mu^\smallFSSM = 251\times 10^{-11}$, as a
    function of the smuon mass and for different values of a common
    mass for higgsinos and EW gauginos. Right: Values of the
    loop-corrected muon Yukawa coupling of the FSSM that correspond to
    the solution for $\tan\beta\tan\beta_d$ shown in the left
    plot. The meaning of the lines is the same as in the left plot.}
  \label{fig:tbtbdymu}
\vspace*{-3mm}
\end{center}
\end{figure}

\bigskip

To set the stage for our discussion, in the left plot of
figure~\ref{fig:tbtbdymu} we show the values of $\tan\beta\tan\beta_d
= v_u/v_{d^\prime}$ that result in the desired smuon-higgsino-gaugino
contribution to $a_\mu$, as a function of the common smuon mass
$M_{\tilde \mu}$ and for different values of $M_\chi$. In particular,
the blue, yellow, green and red solid lines correspond to gaugino and
higgsino masses of $1, 2, 3$ and $4$~TeV, respectively, while the
black dashed line corresponds to the choice $M_\chi=M_{\tilde
  \mu}$. The EW gauge couplings entering eqs.~(\ref{eq:damuFSSM}) and
(\ref{eq:epsilon}) are evaluated at the scale $Q=M_{\tilde \mu}$, but
we found qualitatively similar results for any choice of scale in the
few-TeV range. The plot shows that the values of
$\tan\beta\tan\beta_d$ necessary to obtain $\Delta a_\mu^\smallFSSM =
251\times 10^{-11}$ are typically in the few-hundreds range, and can
reach as much as $940$ for the largest considered value of $M_{\tilde
  \mu}$.

In the right plot of figure~\ref{fig:tbtbdymu} we show instead the
values of the loop-corrected muon Yukawa coupling of the FSSM at the
scale $Q=M_{\tilde \mu}$, 
\beq
\hat y^\prime_\mu(M_{\tilde \mu}) ~=~
\frac{g_\mu(M_{\tilde \mu})/(\cos\tilde\beta\cos\beta_d)}{1+
  \epsilon_\ell\,\tan\beta\tan\beta_d}~,
\label{eq:ymuhat}
\eeq
that correspond to the solutions for $\tan\beta\tan\beta_d$ shown in
the left plot. The FSSM coupling $\hat y^\prime_\mu$ controls the
higgsino-muon-smuon and Higgs-smuon-smuon interactions involved in the
smuon-higgsino-gaugino contributions to $a_\mu$, as well as the
Higgs-smuon-smuon interaction involved in the smuon contribution to
the quartic Higgs coupling. The solid and dashed lines have the same
meaning as in the left plot.  We set $\tan\beta_u=100$ in order to
determine the angle $\tilde\beta$ entering eq.~(\ref{eq:ymuhat}), but
the results are essentially independent of this choice as long as
$\tan\beta_u \gg 1$. The plot shows that larger values of $\hat
y^\prime_\mu$ are required to obtain the desired contribution to
$a_\mu$ when the smuon mass increases, due to the $M_{\tilde
  \mu}^{-2}$ suppression in eq.~(\ref{eq:damuFSSM}). Also, for a fixed
value of the smuon mass, larger values of $\hat y^\prime_\mu$ are
required when $M_\chi$ increases.  We remark that, while the
considered values of $\hat y^\prime_\mu(M_{\tilde \mu})$ are all
perturbative, further constraints on this scenario could arise if we
required that $\hat y^\prime_\mu$ remain perturbative all the way up
to the GUT scale.

\bigskip

In the presence of a large muon Yukawa coupling, the smuon
contribution to the quartic Higgs coupling in eq.~(\ref{eq:dlsmu}) is
dominated by a negative term
\beq
\label{eq:dlsmuapprox}
(4\pi)^2\,\Delta\lambda^{\tilde \mu} ~\approx~
- \frac{\hat y_\mu^{\prime\,4}}{6}\,\left(\frac{M_\chi}{M_{\tilde \mu}}\right)^4,
\eeq
and can become substantial. An increased positive contribution from
loops involving the remaining SUSY particles is then needed to satisfy
the constraint arising from the measured value of the Higgs mass. As
mentioned in the previous section, such positive contribution can most
easily come from the stops, which themselves have generally large
couplings to the SM-like Higgs boson.

\begin{figure}[t]
\begin{center}
\vspace*{-1.2cm} 
\includegraphics[width=12cm]{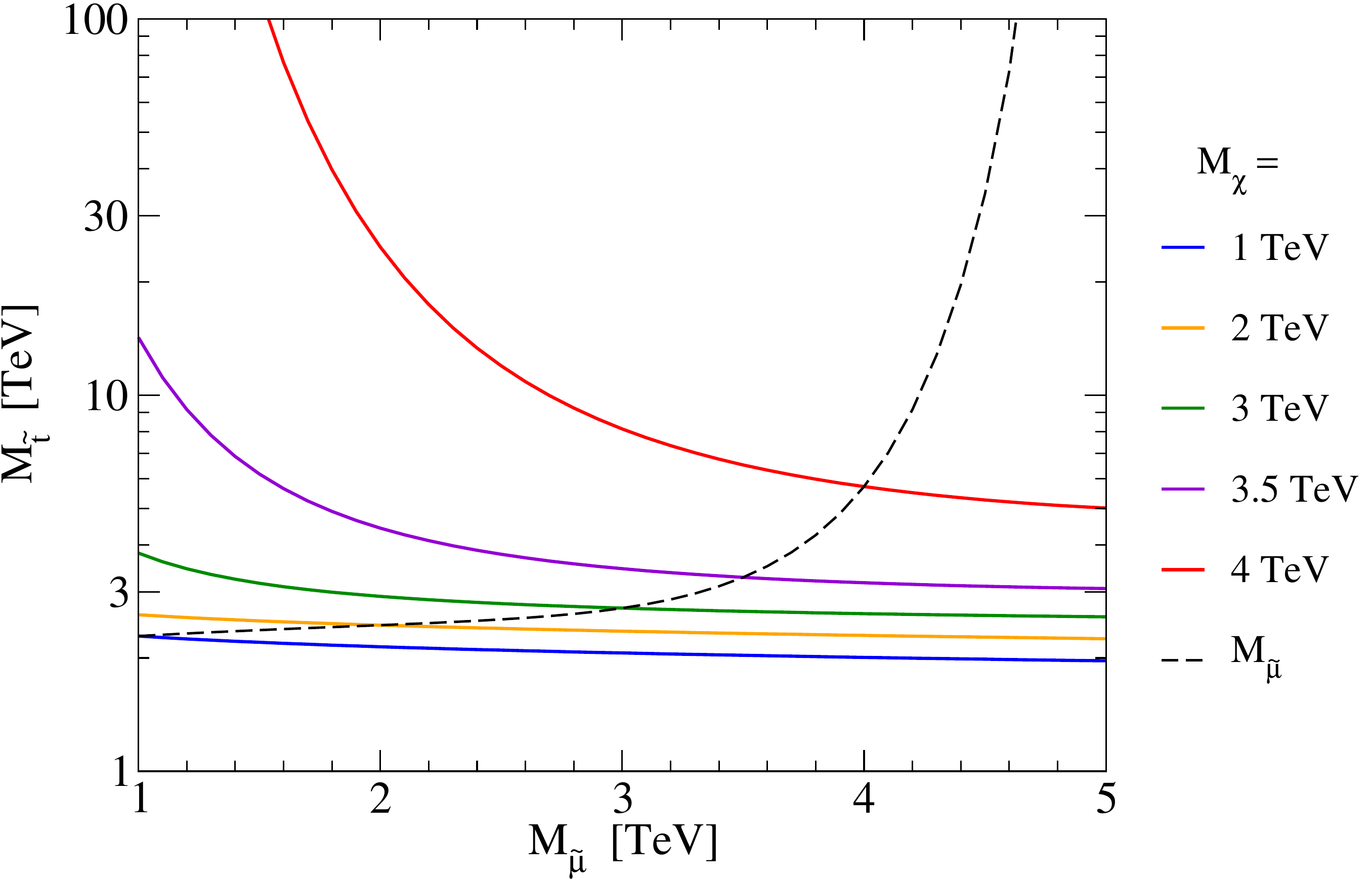}
\vspace*{-1mm}
  \caption{\em Values of the stop mass $M_{\tilde t}$ that result in
    the correct prediction for the Higgs mass when $\tan\beta_d$ is
    fixed by the requirement that the smuon-higgsino-gaugino
    contribution solve the $\gm2$ anomaly, as a function of the smuon
    mass $M_{\tilde \mu}$ and for different values of the common
    higgsino/gaugino mass $M_{\chi}$. The remaining free parameters of
    our simplified FSSM scenario are fixed as $\tan\tilde\beta=20$,
    $\tan\beta_u = 100$, $X_t=\sqrt6\,M_{\tilde t}$ and
    $M_H=M_3=M_\chi$.
    %The matching of the quartic Higgs coupling is performed at the
    %scale $Q=M_{\tilde t}$.
  }
  \label{fig:MstopMsmu}
\vspace*{-5mm}
\end{center}
\end{figure}

In figure~\ref{fig:MstopMsmu} we plot the values of the common stop
mass $M_{\tilde t}$ that are required to obtain the correct prediction
for the quartic Higgs coupling in FSSM scenarios where the
smuon-higgsino-gaugino loops provide the desired contribution to
$\gm2$.  We set the matching scale $Q$ in eqs.~(\ref{eq:lambda1loop})
and (\ref{eq:dlsfe})--(\ref{eq:dlchi}) equal to the stop mass, as this
ensures a full NNL resummation of the large logarithmic corrections
controlled by the top Yukawa coupling. All of the running couplings
entering our calculation are then computed at the scale $Q=M_{\tilde
  t}$.
The common mass parameters for the smuons, $M_{\tilde\mu}$, and for
higgsinos and EW gauginos, $M_\chi$, are varied as in
figure~\ref{fig:tbtbdymu} (note that we add a purple solid line for
$M_\chi = 3.5$~TeV). We fix $\tan\tilde\beta = 20$, thus ensuring that
the tree-level prediction for $\lambdaSM$ in
eq.~(\ref{eq:lambda1loop}) is essentially saturated, and
$\tan\beta_u=100$ (the latter choice has little impact on our results
as long as $\tan\beta_u\gg 1$). In contrast, $\tan\beta_d$ is computed
in each point from the requirement that $\Delta a_\mu^\smallFSSM =
251\times 10^{-11}$.
The stop mixing parameter is fixed to the value $X_t =
\sqrt6\,M_{\tilde t}\,$ that maximizes the one-loop stop contribution
to the quartic Higgs coupling, see eq.~(\ref{eq:dlstop}). Our choices
for $\tan\tilde\beta$ and $X_t$ ensure that the values of the stop
mass shown in figure~\ref{fig:MstopMsmu} are about the minimal ones
that provide the correct prediction for the Higgs mass (in other
words, different choices for these parameters would result in an
overall upward shift of all lines in the figure). For the remaining
free parameters of our simplified FSSM scenario we choose $M_H = M_3 =
M_\chi$. The choice of the common mass parameter for the heavy Higgs
doublets has only a small impact on our results, because the
corresponding contributions to the quartic Higgs coupling, see
eq.~(\ref{eq:dlhig}), all involve four powers of the EW gauge
couplings. The choice of the gluino mass affects our calculation only
through the corrections to the quark Yukawa couplings in
eqs.~(\ref{eq:gqhat}) and (\ref{eq:deltag}), and its qualitative
impact on our results is generally not substantial. Our choice of a
positive value for the ratio $X_t/M_3$ enhances the loop-corrected top
Yukawa coupling $\hat y_t$, whereas a negative value would suppress it
and require somewhat heavier stops in order to satisfy the Higgs-mass
constraint. However, we recall that, for positive values of $X_s/M_3$
(i.e., negative values of $\mu_{ud^\prime}/M_3$), a fine-tuned choice
of parameters such that $\Delta g_s\simeq 1$ might lead the strange
Yukawa coupling to blow up, resulting in a large negative
strange-squark contribution to the quartic Higgs coupling.

Figure~\ref{fig:MstopMsmu} shows that there are regions of the FSSM
parameter space in which the interplay between the requirements of a
suitable contribution to $a_\mu$ and of a correct prediction for the
quartic Higgs coupling implies a strongly hierarchical SUSY spectrum,
with the stops being significantly heavier than smuons, higgsinos and
EW gauginos. Unsurprisingly, in the scenario with $M_\chi =
M_{\tilde\mu}$ (the black dashed line) this happens for the largest
values of $M_{\tilde\mu}$, when a large muon Yukawa coupling $\hat
y^\prime_\mu$ is needed to counteract the $M_{\tilde\mu}^{-2}$
suppression of the smuon-higgsino-gaugino contribution to $a_\mu$, see
eq.~(\ref{eq:damuFSSM}), and results in a large negative contribution
to the quartic Higgs coupling, see
eq.~(\ref{eq:dlsmuapprox}). Moreover, the left ends of the red, purple
and (to a lesser extent) green lines show that very heavy stops may be
needed also at lower values of $M_{\tilde\mu}$ -- where
figure~\ref{fig:tbtbdymu} shows that $\hat y^\prime_\mu \lesssim 1$ is
sufficient for the solution of the $\gm2$ anomaly -- if the smuon
contribution to the quartic Higgs coupling in
eq.~(\ref{eq:dlsmuapprox}) is enhanced by a large ratio
$M_\chi/M_{\tilde\mu}$.
Even in regions of the parameter space where the SUSY spectrum is not
strongly hierarchical, such as the right end of the red line in
figure~\ref{fig:MstopMsmu}, the minimal value of $M_{\tilde t}$
required to obtain the correct Higgs-mass prediction can be
significantly higher than the $2\!-\!3$~TeV typically found in the
MSSM.
Finally, we find that for $M_\chi \lesssim 2$~TeV (the yellow and blue
lines) the smuon contribution to the quartic Higgs coupling never
becomes substantial in our simplified scenario: at low $M_{\tilde
  \mu}$ -- where the desired contribution to $\gm2$ is obtained with
$\hat y^\prime_\mu\lesssim 0.5$ -- there is little or no enhancement
from the $(M_\chi/M_{\tilde \mu})^4$ term in
eq.~(\ref{eq:dlsmuapprox}), whereas at high $M_{\tilde \mu}$ the
corresponding suppression wins over the enhancement of $\hat
y^\prime_\mu$. The mild residual dependence of the yellow and blue
lines on $M_{\tilde \mu}$ and $M_\chi$ stems from the terms controlled
by the EW gauge couplings in eqs.~(\ref{eq:dlsfe})--(\ref{eq:dlchi}).

We remark that in our simplified FSSM scenario, where $M_\chi \equiv
M_1=M_2=\mu=\tilde\mu$\,, the condition $M_\chi/M_{\tilde\mu}>1$
implies that the LSP is a heavy sneutrino, which is generally
disfavored by Dark Matter
considerations~\cite{Falk:1994es,Arina:2007tm}. While a detailed study
of Dark Matter constraints on the FSSM parameter space is beyond the
scope of our paper, using the general formula for $\Delta\lambda^\chi$
we verified that very heavy stops may be required even in scenarios
where the LSP is always an EW gaugino. In particular, if we fix $M_1 =
1$~TeV and $M_\chi \equiv M_2=\mu=\tilde\mu$ we still find a rise in
the stop mass similar to the black dashed line in
figure~\ref{fig:MstopMsmu} for $M_{\tilde \mu} = M_\chi \,\gtrsim
\,4$~TeV. However, we no longer see the rise at low $M_{\tilde \mu}$
in the lines corresponding to larger $M_\chi$, because in this region
the desired contribution to $\gm2$ is obtained with lower values of
$\hat y^\prime_\mu$ than in the case of degenerate gaugino masses.

\bigskip

It is interesting to note that, in contrast to what we find for the
FSSM, in the MSSM the combination of the constraints from the
Higgs-mass prediction and from $\gm2$ can yield {\em upper} bounds on
the stop masses, see e.g.~ref.~\cite{Badziak:2014kea}. Indeed, in the
absence of large and negative contributions from other sectors, a
large and positive contribution from heavy-stop loops to the
prediction for $\lambdaSM$ must be compensated for by a suppression of
the tree-level prediction via a lower value of $\tan\beta$. However,
in the MSSM this would also suppress the smuon-higgsino-gaugino
contribution to $a_\mu$, jeopardizing the solution of the $\gm2$
anomaly. The interplay of the two constraints is different in the
FSSM, because in this model the ratio of vevs that determines the
tree-level prediction for $\lambdaSM$ can be varied independently of
the ratio of vevs that enhances the smuon-higgsino-gaugino
contribution to $a_\mu$.

\bigskip

Finally, we remark that, when the stops are an order of magnitude (or
more) heavier than the other SUSY particles, our calculation of the
Higgs mass loses accuracy, and we would need to build a two-step EFT
setup in which the stops are separately integrated out of the FSSM at
a scale comparable with their mass. However, the aim of our study is
not a precise determination of masses that, for the time being, are
beyond experimental reach, but rather a qualitative insight on the
structure of this heavy-SUSY model.\footnote{For the same reason we do
not take into account the theoretical uncertainties of our predictions
for $\lambdaSM$ and $a_\mu$.} The possibility of a hierarchical mass
spectrum has obvious implications for the prospects of probing the
FSSM at future colliders, and from the model-building point of view it
might also complicate any attempt to devise a suitable mechanism of
SUSY breaking.

\vfill
\newpage

\section{Conclusions}
\label{sec:conclusions}

A scenario for particle physics that is now looking increasingly
plausible is the one where new physics manifests itself in one or more
deviations from the SM predictions for rare processes or precision
observables, but the BSM particles responsible for those deviations
are too heavy to be discovered at the LHC. In this case, all possible
clues should be exploited to unravel the structure of the heavy BSM
sector, also to guide the searches for the new particles at future
colliders.
If a model that aims to explain the observed anomalies is
supersymmetric, it will generally involve a prediction for the quartic
coupling $\lambdaSM$ of the SM-like Higgs boson. Since all of the new
particles in the SUSY model affect $\lambdaSM$ through radiative
corrections, its prediction can reveal correlations between the
sectors of the model that are involved in the observed anomalies and
those that are not.

In this paper we studied the Higgs-mass constraints on the parameter
space of a supersymmetric four-Higgs-doublet model, the
FSSM~\cite{Altmannshofer:2021hfu}, which was recently proposed as a
solution of the $\gm2$ anomaly~\cite{Muong-2:2021ojo, Muong-2:2006rrc}
with SUSY particle masses beyond the current reach of the LHC. We
followed the modern approach of taking $M_h$ as an input rather than
an output of our calculation, and we relied on an EFT setup to account
at the NLL order for the large logarithmic corrections to the relation
between the measured value of $M_h$ and the value of $\lambdaSM$ at
the SUSY scale. In our one-loop calculation of the prediction for
$\lambdaSM$ we adapted to the case of the FSSM the results derived in
ref.~\cite{Braathen:2018htl} for a general renormalizable theory. We
provided explicit formulas for the one-loop correction to the quartic
Higgs coupling in a simplified FSSM scenario, but we make the result
for the general FSSM available on request in electronic form.

We found that the prediction for $\lambdaSM$ establishes interesting
relations between the parameters that contribute to $\gm2$, namely the
masses of smuons, higgsinos and EW gauginos, and the parameters in the
stop sector. In particular, there are scenarios with a suitable SUSY
contribution to $\gm2$ in which the stops need to be considerably
heavier than smuons, higgsinos and EW gauginos, in order to compensate
for a large and negative smuon contribution to the prediction for
$\lambdaSM$. The possibility of a hierarchical SUSY mass spectrum
should be taken into account when assessing the prospects of probing
the FSSM at future colliders. 

As mentioned in ref.~\cite{Altmannshofer:2021hfu}, further
investigations of the FSSM could address the flavor structure of the
quark sector, in which case the large corrections to the strange
Yukawa coupling discussed in section~\ref{sec:matching} of our paper
would need to be taken into account. Other directions of investigation
could address the extended Higgs sector of the FSSM, exploring
e.g.~the collider phenomenology of the heavy Higgs bosons, or the
stability of the scalar potential. With the present study, we aimed to
provide a proof of concept -- applicable also to other models and
other anomalies -- of how the Higgs-mass prediction can be used, in
combination with other observables, to shed some light on the hidden
structure of a SUSY model with heavy superparticles.

\section*{Acknowledgments}

We thank Karim Benakli and Dominik St\"ockinger for useful
discussions.
The work of W.~K.~is supported by the Contrat Doctoral Sp\'ecifique
Normalien (CDSN) of Ecole Normale Sup\'erieure -- PSL.
The work of P.~S.~is supported in part by French state funds managed
by the Agence Nationale de la Recherche (ANR), in the context of the
grant ``HiggsAutomator'' (ANR-15-CE31-0002).

\begin{Appendix}
\section*{Appendix}

We provide here explicit formulas for the tree-level mass matrices of
the Higgs bosons in the FSSM. In the Higgs basis, we decompose the
four SU(2) doublets (all with positive hypercharge) as
\beq
\Phi_h ~=\,
\left(\!\begin{array}{c} G^+ \\ v + \frac1{\sqrt2} (h + i \,G^0)
\end{array}\!\right),~~~~~
\Phi_H ~=\,
\left(\!\begin{array}{c} H^+ \\ \frac1{\sqrt2} (H + i  \,A)
\end{array}\!\right),~~~~~
\Phi^\prime_{d,u} ~=\,
\left(\!\begin{array}{c} \phi^{\prime\,+}_{d,u} \\
\frac1{\sqrt2} ( \phi^{\prime}_{d,u} + i  \,a^{\prime}_{d,u})
\end{array}\!\right).
\eeq
In this basis the mass matrices for the scalar
$(h,\,H,\,\phi^{\prime}_{d},\,\phi^{\prime}_{u})$, pseudoscalar
$(G^0,\,A,\,a^{\prime}_{d},\,a^{\prime}_{u})$ and charged
$(G^+,\,H^+,\,\phi^{\prime\,+}_{d},\,\phi^{\prime\,+}_{u})$ components
of the four doublets read

\beq
{\cal M}^2_S ~=~
\left(\!\begin{array}{cccc}
\MZ^2\,\cos^22\tilde\beta & -\frac12\,\MZ^2\,\sin4\tilde\beta & 0 & 0\\[1mm]
-\frac12\,\MZ^2\,\sin4\tilde\beta &
\MZ^2\,\sin^22\tilde\beta + 2 \,b_{12}/\sin2\tilde\beta&
-b_{32}/\cos\tilde\beta & b_{14}/\sin\tilde\beta\\[1mm]
0 & -b_{32}/\cos\tilde\beta &
M_{\Phi_d^\prime}^2 + \frac12\,\MZ^2\,\cos2\tilde\beta & - b_{34}\\[1mm]
0 & b_{14}/\sin\tilde\beta & - b_{34} &
M_{\Phi_u^\prime}^2 - \frac12\,\MZ^2\,\cos2\tilde\beta
\end{array}\!\right),
\label{eq:MS}
\eeq

\beq
{\cal M}^2_P ~=~
\left(\!\begin{array}{cccc}
0 & 0 & 0 & 0\\[1mm]
0 & 2 \,b_{12}/\sin2\tilde\beta&
-b_{32}/\cos\tilde\beta & b_{14}/\sin\tilde\beta\\[1mm]
0 & -b_{32}/\cos\tilde\beta &
M_{\Phi_d^\prime}^2 + \frac12\,\MZ^2\,\cos2\tilde\beta & - b_{34}\\[1mm]
0 & b_{14}/\sin\tilde\beta & - b_{34} &
M_{\Phi_u^\prime}^2 - \frac12\,\MZ^2\,\cos2\tilde\beta
\end{array}\!\right),
\label{eq:MP}
\eeq

\beq
{\cal M}^2_\pm ~=~
\left(\!\begin{array}{cccc}
0 & 0 & 0 & 0\\[1mm]
0 & \MW^2 + 2 \,b_{12}/\sin2\tilde\beta&
-b_{32}/\cos\tilde\beta & b_{14}/\sin\tilde\beta\\[1mm]
0 & -b_{32}/\cos\tilde\beta &
M_{\Phi_d^\prime}^2 - \left(\MW^2 -\frac12\,\MZ^2\right)\,\cos2\tilde\beta
& - b_{34}\\[1mm]
0 & b_{14}/\sin\tilde\beta & - b_{34} &
M_{\Phi_u^\prime}^2 +\left(\MW^2- \frac12\,\MZ^2\right)\,\cos2\tilde\beta
\end{array}\!\right),
\label{eq:MC}
\eeq

\noindent where the mass parameters $M_{\Phi_{d,u}^\prime}^2$ and the
mixing parameters\,\footnote{Our notation for the mixing parameters
$b_{ij}$ follows ref.~\cite{Escudero:2005hk}. Note however that the
upper-left $4\!\times\!4$ blocks of the mass matrices shown in
eqs.~(31), (34) and (35) of that paper correspond to a different
basis, namely $\left(\epsilon \Phi^*_d,\,\Phi_u,\, \epsilon
\Phi^{\prime\,*}_d,\,\Phi^\prime_u\right)$.} $b_{ij}$ are combinations
of the original mass parameters and $B$-terms as defined in
eqs.~(\ref{eq:Wmu}) and (\ref{eq:LsoftH}):

\bea
M_{\Phi_d^\prime}^2 &=& 
\left(m_{dd}^2+\mu_{ud}^2+\mu_{u^\prime \!d}^2\right)\cos^2\beta_d
~+~\left(m_{d^\prime\!d^\prime}^2+\mu_{u^\prime \!d^\prime}^2+\mu_{u d^\prime}^2
\right)\sin^2\beta_d
\nn\\[1mm]
&& - \left(m^2_{dd^\prime} + \mu_{ud}\mu_{u d^\prime}+ \mu_{u^\prime \!d^\prime}\mu_{u^\prime \!d}
\right)\sin2\beta_d~,\\[3mm]
M_{\Phi_u^\prime}^2 &=& 
\left(m_{uu}^2+\mu_{ud}^2+\mu_{u d^\prime}^2\right)\cos^2\beta_u
~+~\left(m_{u^\prime\!u^\prime}^2+\mu_{u^\prime \!d^\prime}^2+\mu_{u^\prime \!d}^2
\right)\sin^2\beta_u
\nn\\[1mm]
&& - \left(m^2_{uu^\prime} + \mu_{ud}\mu_{u^\prime \!d}+ \mu_{u^\prime \!d^\prime}\mu_{u d^\prime}
\right)\sin2\beta_u~,
\eea
\bea
b_{12} &=& \sin\beta_d\,\left(B_{ud}\sin\beta_u+B_{u^\prime \!d}\cos\beta_u\right)
~+~\cos\beta_d\,\left(B_{ud^\prime}\sin\beta_u+B_{u^\prime \!d^\prime}\cos\beta_u
\right)~,\\[2mm]
b_{32} &=& \cos\beta_d\,\left(B_{ud}\sin\beta_u+B_{u^\prime \!d}\cos\beta_u\right)
~-~\sin\beta_d\,\left(B_{ud^\prime}\sin\beta_u+B_{u^\prime \!d^\prime}\cos\beta_u
\right)~,\\[2mm]
b_{14} &=& \sin\beta_d\,\left(B_{ud}\cos\beta_u-B_{u^\prime \!d}\sin\beta_u\right)
~+~\cos\beta_d\,\left(B_{ud^\prime}\cos\beta_u-B_{u^\prime \!d^\prime}\sin\beta_u
\right)~,\\[2mm]
b_{34} &=& \cos\beta_d\,\left(B_{ud}\cos\beta_u-B_{u^\prime \!d}\sin\beta_u\right)
~-~\sin\beta_d\,\left(B_{ud^\prime}\cos\beta_u-B_{u^\prime \!d^\prime}\sin\beta_u
\right)~.
\eea

The minimum conditions of the Higgs potential have been used to remove
four combinations of the original parameters from
eqs.~(\ref{eq:MS})--(\ref{eq:MC}). In the limit of unbroken EW
symmetry (i.e., $v\rightarrow 0$), which we adopt in the calculation
of the matching condition for the quartic coupling of the SM-like
Higgs, the mixing between the SM-like scalar $h$ and the three heavy
scalars vanishes, and the $3\!\times\!3$ sub-matrices for the masses
of the scalar, pseudoscalar and charged components of the heavy
doublets $\Phi_H$, $\Phi^\prime_{d}$ and $\Phi^\prime_{u}$ all reduce
to:

\beq
{\cal M}^2_H ~=~
\left(\!\begin{array}{ccc}
2 \,b_{12}/\sin2\tilde\beta&
-b_{32}/\cos\tilde\beta & b_{14}/\sin\tilde\beta\\[1mm]
-b_{32}/\cos\tilde\beta &
M_{\Phi_d^\prime}^2 & - b_{34}\\[1mm]
b_{14}/\sin\tilde\beta & - b_{34} &
M_{\Phi_u^\prime}^2
\end{array}\!\right).
\label{eq:MH}
\eeq

\bigskip
\noindent We can then introduce a $3\!\times\!3$ orthogonal matrix
$R_H$ that rotates the three heavy doublets of the Higgs basis into a
basis of mass eigenstates:

\beq
\left(\!\begin{array}{c} H_1 \\ H_2 \\ H_3
\end{array}\!\right) ~=~
R_H\,\left(\!\begin{array}{c} \Phi_\smallH \\ \Phi_d^\prime \\ \Phi_u^\prime
\end{array}\!\right)~, ~~~~~~~~~~~
{\rm diag}\left(M_{H_1}^2,M_{H_2}^2,M_{H_3}^2\right) ~=~ R_H \, {\cal M}^2_H\,R_H^T~.
\eeq

\end{Appendix}
\vfill
\newpage
\bibliographystyle{utphys}
\bibliography{KS}

\providecommand{\href}[2]{#2}\begingroup\raggedright\begin{thebibliography}{10}

\bibitem{CMS:2012qbp}
{\bf CMS} Collaboration, S.~Chatrchyan {\em et al.}, {\em {Observation of a New
  Boson at a Mass of 125 GeV with the CMS Experiment at the LHC}}.
  \href{http://dx.doi.org/10.1016/j.physletb.2012.08.021}{Phys. Lett. B {\bf
  716} (2012)  30--61}, \href{http://arxiv.org/abs/1207.7235}{{\tt
  arXiv:1207.7235 [hep-ex]}}.

\bibitem{ATLAS:2012yve}
{\bf ATLAS} Collaboration, G.~Aad {\em et al.}, {\em {Observation of a new
  particle in the search for the Standard Model Higgs boson with the ATLAS
  detector at the LHC}}.
  \href{http://dx.doi.org/10.1016/j.physletb.2012.08.020}{Phys. Lett. B {\bf
  716} (2012)  1--29}, \href{http://arxiv.org/abs/1207.7214}{{\tt
  arXiv:1207.7214 [hep-ex]}}.

\bibitem{ATLAS:2015yey}
{\bf ATLAS, CMS} Collaboration, G.~Aad {\em et al.}, {\em {Combined Measurement
  of the Higgs Boson Mass in $pp$ Collisions at $\sqrt{s}=7$ and 8 TeV with the
  ATLAS and CMS Experiments}}.
  \href{http://dx.doi.org/10.1103/PhysRevLett.114.191803}{Phys. Rev. Lett. {\bf
  114} (2015)  191803}, \href{http://arxiv.org/abs/1503.07589}{{\tt
  arXiv:1503.07589 [hep-ex]}}.

\bibitem{ATLAS:2016neq}
{\bf ATLAS, CMS} Collaboration, G.~Aad {\em et al.}, {\em {Measurements of the
  Higgs boson production and decay rates and constraints on its couplings from
  a combined ATLAS and CMS analysis of the LHC pp collision data at $
  \sqrt{s}=7 $ and 8 TeV}}.
  \href{http://dx.doi.org/10.1007/JHEP08(2016)045}{JHEP {\bf 08} (2016)  045},
  \href{http://arxiv.org/abs/1606.02266}{{\tt arXiv:1606.02266 [hep-ex]}}.

\bibitem{Slavich:2020zjv}
P.~Slavich {\em et al.}, {\em {Higgs-mass predictions in the MSSM and beyond}}.
  \href{http://dx.doi.org/10.1140/epjc/s10052-021-09198-2}{Eur. Phys. J. C {\bf
  81} (2021) no.~5, 450}, \href{http://arxiv.org/abs/2012.15629}{{\tt
  arXiv:2012.15629 [hep-ph]}}.

\bibitem{LHCb:2014vgu}
{\bf LHCb} Collaboration, R.~Aaij {\em et al.}, {\em {Test of lepton
  universality using $B^{+}\rightarrow K^{+}\ell^{+}\ell^{-}$ decays}}.
  \href{http://dx.doi.org/10.1103/PhysRevLett.113.151601}{Phys. Rev. Lett. {\bf
  113} (2014)  151601}, \href{http://arxiv.org/abs/1406.6482}{{\tt
  arXiv:1406.6482 [hep-ex]}}.

\bibitem{LHCb:2017avl}
{\bf LHCb} Collaboration, R.~Aaij {\em et al.}, {\em {Test of lepton
  universality with $B^{0} \rightarrow K^{*0}\ell^{+}\ell^{-}$ decays}}.
  \href{http://dx.doi.org/10.1007/JHEP08(2017)055}{JHEP {\bf 08} (2017)  055},
  \href{http://arxiv.org/abs/1705.05802}{{\tt arXiv:1705.05802 [hep-ex]}}.

\bibitem{LHCb:2019hip}
{\bf LHCb} Collaboration, R.~Aaij {\em et al.}, {\em {Search for
  lepton-universality violation in $B^+\to K^+\ell^+\ell^-$ decays}}.
  \href{http://dx.doi.org/10.1103/PhysRevLett.122.191801}{Phys. Rev. Lett. {\bf
  122} (2019) no.~19, 191801}, \href{http://arxiv.org/abs/1903.09252}{{\tt
  arXiv:1903.09252 [hep-ex]}}.

\bibitem{LHCb:2021trn}
{\bf LHCb} Collaboration, R.~Aaij {\em et al.}, {\em {Test of lepton
  universality in beauty-quark decays}}.
  \href{http://arxiv.org/abs/2103.11769}{{\tt arXiv:2103.11769 [hep-ex]}}.

\bibitem{Muong-2:2021ojo}
{\bf Muon g-2} Collaboration, B.~Abi {\em et al.}, {\em {Measurement of the
  Positive Muon Anomalous Magnetic Moment to 0.46 ppm}}.
  \href{http://dx.doi.org/10.1103/PhysRevLett.126.141801}{Phys. Rev. Lett. {\bf
  126} (2021) no.~14, 141801}, \href{http://arxiv.org/abs/2104.03281}{{\tt
  arXiv:2104.03281 [hep-ex]}}.

\bibitem{Muong-2:2006rrc}
{\bf Muon g-2} Collaboration, G.~W. Bennett {\em et al.}, {\em {Final Report of
  the Muon E821 Anomalous Magnetic Moment Measurement at BNL}}.
  \href{http://dx.doi.org/10.1103/PhysRevD.73.072003}{Phys. Rev. D {\bf 73}
  (2006)  072003}, \href{http://arxiv.org/abs/hep-ex/0602035}{{\tt
  arXiv:hep-ex/0602035}}.

\bibitem{Aoyama:2020ynm}
T.~Aoyama {\em et al.}, {\em {The anomalous magnetic moment of the muon in the
  Standard Model}}.
  \href{http://dx.doi.org/10.1016/j.physrep.2020.07.006}{Phys. Rept. {\bf 887}
  (2020)  1--166}, \href{http://arxiv.org/abs/2006.04822}{{\tt arXiv:2006.04822
  [hep-ph]}}.

\bibitem{Aoyama:2012wk}
T.~Aoyama, M.~Hayakawa, T.~Kinoshita, and M.~Nio, {\em {Complete Tenth-Order
  QED Contribution to the Muon g-2}}.
  \href{http://dx.doi.org/10.1103/PhysRevLett.109.111808}{Phys. Rev. Lett. {\bf
  109} (2012)  111808}, \href{http://arxiv.org/abs/1205.5370}{{\tt
  arXiv:1205.5370 [hep-ph]}}.

\bibitem{Aoyama:2019ryr}
T.~Aoyama, T.~Kinoshita, and M.~Nio, {\em {Theory of the Anomalous Magnetic
  Moment of the Electron}}. \href{http://dx.doi.org/10.3390/atoms7010028}{Atoms
  {\bf 7} (2019) no.~1, 28}.

\bibitem{Czarnecki:2002nt}
A.~Czarnecki, W.~J. Marciano, and A.~Vainshtein, {\em {Refinements in
  electroweak contributions to the muon anomalous magnetic moment}}.
  \href{http://dx.doi.org/10.1103/PhysRevD.67.073006}{Phys. Rev. D {\bf 67}
  (2003)  073006}, \href{http://arxiv.org/abs/hep-ph/0212229}{{\tt
  arXiv:hep-ph/0212229}}. [Erratum: Phys.Rev.D 73, 119901 (2006)].

\bibitem{Gnendiger:2013pva}
C.~Gnendiger, D.~St\"ockinger, and H.~St\"ockinger-Kim, {\em {The electroweak
  contributions to $(g-2)_\mu$ after the Higgs boson mass measurement}}.
  \href{http://dx.doi.org/10.1103/PhysRevD.88.053005}{Phys. Rev. D {\bf 88}
  (2013)  053005}, \href{http://arxiv.org/abs/1306.5546}{{\tt arXiv:1306.5546
  [hep-ph]}}.

\bibitem{Davier:2017zfy}
M.~Davier, A.~Hoecker, B.~Malaescu, and Z.~Zhang, {\em {Reevaluation of the
  hadronic vacuum polarisation contributions to the Standard Model predictions
  of the muon $g-2$ and ${\alpha (m_Z^2)}$ using newest hadronic cross-section
  data}}. \href{http://dx.doi.org/10.1140/epjc/s10052-017-5161-6}{Eur. Phys. J.
  C {\bf 77} (2017) no.~12, 827}, \href{http://arxiv.org/abs/1706.09436}{{\tt
  arXiv:1706.09436 [hep-ph]}}.

\bibitem{Keshavarzi:2018mgv}
A.~Keshavarzi, D.~Nomura, and T.~Teubner, {\em {Muon $g-2$ and $\alpha(M_Z^2)$:
  a new data-based analysis}}.
  \href{http://dx.doi.org/10.1103/PhysRevD.97.114025}{Phys. Rev. D {\bf 97}
  (2018) no.~11, 114025}, \href{http://arxiv.org/abs/1802.02995}{{\tt
  arXiv:1802.02995 [hep-ph]}}.

\bibitem{Colangelo:2018mtw}
G.~Colangelo, M.~Hoferichter, and P.~Stoffer, {\em {Two-pion contribution to
  hadronic vacuum polarization}}.
  \href{http://dx.doi.org/10.1007/JHEP02(2019)006}{JHEP {\bf 02} (2019)  006},
  \href{http://arxiv.org/abs/1810.00007}{{\tt arXiv:1810.00007 [hep-ph]}}.

\bibitem{Hoferichter:2019gzf}
M.~Hoferichter, B.-L. Hoid, and B.~Kubis, {\em {Three-pion contribution to
  hadronic vacuum polarization}}.
  \href{http://dx.doi.org/10.1007/JHEP08(2019)137}{JHEP {\bf 08} (2019)  137},
  \href{http://arxiv.org/abs/1907.01556}{{\tt arXiv:1907.01556 [hep-ph]}}.

\bibitem{Davier:2019can}
M.~Davier, A.~Hoecker, B.~Malaescu, and Z.~Zhang, {\em {A new evaluation of the
  hadronic vacuum polarisation contributions to the muon anomalous magnetic
  moment and to $\mathbf{\boldsymbol\alpha(m_Z^2)}$}}.
  \href{http://dx.doi.org/10.1140/epjc/s10052-020-7792-2}{Eur. Phys. J. C {\bf
  80} (2020) no.~3, 241}, \href{http://arxiv.org/abs/1908.00921}{{\tt
  arXiv:1908.00921 [hep-ph]}}. [Erratum: Eur.Phys.J.C 80, 410 (2020)].

\bibitem{Keshavarzi:2019abf}
A.~Keshavarzi, D.~Nomura, and T.~Teubner, {\em {$g-2$ of charged leptons,
  $\alpha (M^2_Z)$ , and the hyperfine splitting of muonium}}.
  \href{http://dx.doi.org/10.1103/PhysRevD.101.014029}{Phys. Rev. D {\bf 101}
  (2020) no.~1, 014029}, \href{http://arxiv.org/abs/1911.00367}{{\tt
  arXiv:1911.00367 [hep-ph]}}.

\bibitem{Kurz:2014wya}
A.~Kurz, T.~Liu, P.~Marquard, and M.~Steinhauser, {\em {Hadronic contribution
  to the muon anomalous magnetic moment to next-to-next-to-leading order}}.
  \href{http://dx.doi.org/10.1016/j.physletb.2014.05.043}{Phys. Lett. B {\bf
  734} (2014)  144--147}, \href{http://arxiv.org/abs/1403.6400}{{\tt
  arXiv:1403.6400 [hep-ph]}}.

\bibitem{Melnikov:2003xd}
K.~Melnikov and A.~Vainshtein, {\em {Hadronic light-by-light scattering
  contribution to the muon anomalous magnetic moment revisited}}.
  \href{http://dx.doi.org/10.1103/PhysRevD.70.113006}{Phys. Rev. D {\bf 70}
  (2004)  113006}, \href{http://arxiv.org/abs/hep-ph/0312226}{{\tt
  arXiv:hep-ph/0312226}}.

\bibitem{Masjuan:2017tvw}
P.~Masjuan and P.~Sanchez-Puertas, {\em {Pseudoscalar-pole contribution to the
  $(g_{\mu}-2)$: a rational approach}}.
  \href{http://dx.doi.org/10.1103/PhysRevD.95.054026}{Phys. Rev. D {\bf 95}
  (2017) no.~5, 054026}, \href{http://arxiv.org/abs/1701.05829}{{\tt
  arXiv:1701.05829 [hep-ph]}}.

\bibitem{Colangelo:2017fiz}
G.~Colangelo, M.~Hoferichter, M.~Procura, and P.~Stoffer, {\em {Dispersion
  relation for hadronic light-by-light scattering: two-pion contributions}}.
  \href{http://dx.doi.org/10.1007/JHEP04(2017)161}{JHEP {\bf 04} (2017)  161},
  \href{http://arxiv.org/abs/1702.07347}{{\tt arXiv:1702.07347 [hep-ph]}}.

\bibitem{Hoferichter:2018kwz}
M.~Hoferichter, B.-L. Hoid, B.~Kubis, S.~Leupold, and S.~P. Schneider, {\em
  {Dispersion relation for hadronic light-by-light scattering: pion pole}}.
  \href{http://dx.doi.org/10.1007/JHEP10(2018)141}{JHEP {\bf 10} (2018)  141},
  \href{http://arxiv.org/abs/1808.04823}{{\tt arXiv:1808.04823 [hep-ph]}}.

\bibitem{Gerardin:2019vio}
A.~G\'erardin, H.~B. Meyer, and A.~Nyffeler, {\em {Lattice calculation of the
  pion transition form factor with $N_f=2+1$ Wilson quarks}}.
  \href{http://dx.doi.org/10.1103/PhysRevD.100.034520}{Phys. Rev. D {\bf 100}
  (2019) no.~3, 034520}, \href{http://arxiv.org/abs/1903.09471}{{\tt
  arXiv:1903.09471 [hep-lat]}}.

\bibitem{Bijnens:2019ghy}
J.~Bijnens, N.~Hermansson-Truedsson, and A.~Rodr\'\i{}guez-S\'anchez, {\em
  {Short-distance constraints for the HLbL contribution to the muon anomalous
  magnetic moment}}.
  \href{http://dx.doi.org/10.1016/j.physletb.2019.134994}{Phys. Lett. B {\bf
  798} (2019)  134994}, \href{http://arxiv.org/abs/1908.03331}{{\tt
  arXiv:1908.03331 [hep-ph]}}.

\bibitem{Colangelo:2019uex}
G.~Colangelo, F.~Hagelstein, M.~Hoferichter, L.~Laub, and P.~Stoffer, {\em
  {Longitudinal short-distance constraints for the hadronic light-by-light
  contribution to $(g-2)_\mu$ with large-$N_c$ Regge models}}.
  \href{http://dx.doi.org/10.1007/JHEP03(2020)101}{JHEP {\bf 03} (2020)  101},
  \href{http://arxiv.org/abs/1910.13432}{{\tt arXiv:1910.13432 [hep-ph]}}.

\bibitem{Blum:2019ugy}
T.~Blum, N.~Christ, M.~Hayakawa, T.~Izubuchi, L.~Jin, C.~Jung, and C.~Lehner,
  {\em {Hadronic Light-by-Light Scattering Contribution to the Muon Anomalous
  Magnetic Moment from Lattice QCD}}.
  \href{http://dx.doi.org/10.1103/PhysRevLett.124.132002}{Phys. Rev. Lett. {\bf
  124} (2020) no.~13, 132002}, \href{http://arxiv.org/abs/1911.08123}{{\tt
  arXiv:1911.08123 [hep-lat]}}.

\bibitem{Colangelo:2014qya}
G.~Colangelo, M.~Hoferichter, A.~Nyffeler, M.~Passera, and P.~Stoffer, {\em
  {Remarks on higher-order hadronic corrections to the muon g\ensuremath{-}2}}.
  \href{http://dx.doi.org/10.1016/j.physletb.2014.06.012}{Phys. Lett. B {\bf
  735} (2014)  90--91}, \href{http://arxiv.org/abs/1403.7512}{{\tt
  arXiv:1403.7512 [hep-ph]}}.

\bibitem{Altmannshofer:2010zt}
W.~Altmannshofer and D.~M. Straub, {\em {Viability of MSSM scenarios at very
  large $\tan\beta$}}. \href{http://dx.doi.org/10.1007/JHEP09(2010)078}{JHEP
  {\bf 09} (2010)  078}, \href{http://arxiv.org/abs/1004.1993}{{\tt
  arXiv:1004.1993 [hep-ph]}}.

\bibitem{Chakraborti:2021dli}
M.~Chakraborti, S.~Heinemeyer, and I.~Saha, {\em {The new
  \textquotedblleft{}MUON G-2\textquotedblright{} result and supersymmetry}}.
  \href{http://dx.doi.org/10.1140/epjc/s10052-021-09900-4}{Eur. Phys. J. C {\bf
  81} (2021) no.~12, 1114}, \href{http://arxiv.org/abs/2104.03287}{{\tt
  arXiv:2104.03287 [hep-ph]}}.

\bibitem{Athron:2021iuf}
P.~Athron, C.~Bal\'azs, D.~H. Jacob, W.~Kotlarski, D.~St\"ockinger, and
  H.~St\"ockinger-Kim, {\em {New physics explanations of $a_\mu$ in light of
  the FNAL muon $g-2$ measurement}}.
  \href{http://dx.doi.org/10.1007/JHEP09(2021)080}{JHEP {\bf 09} (2021)  080},
  \href{http://arxiv.org/abs/2104.03691}{{\tt arXiv:2104.03691 [hep-ph]}}.

\bibitem{Bach:2015doa}
M.~Bach, J.-h. Park, D.~St\"ockinger, and H.~St\"ockinger-Kim, {\em {Large muon
  $(g-2)$ with TeV-scale SUSY masses for $\tan\beta\to\infty$}}.
  \href{http://dx.doi.org/10.1007/JHEP10(2015)026}{JHEP {\bf 10} (2015)  026},
  \href{http://arxiv.org/abs/1504.05500}{{\tt arXiv:1504.05500 [hep-ph]}}.

\bibitem{Altmannshofer:2021hfu}
W.~Altmannshofer, S.~A. Gadam, S.~Gori, and N.~Hamer, {\em {Explaining
  $(g-2)_{\mu}$ with Multi-TeV Sleptons}}.
  \href{http://dx.doi.org/10.1007/JHEP07(2021)118}{JHEP {\bf 07} (2021)  118},
  \href{http://arxiv.org/abs/2104.08293}{{\tt arXiv:2104.08293 [hep-ph]}}.

\bibitem{Benakli:2013msa}
K.~Benakli, L.~Darm\'e, M.~D. Goodsell, and P.~Slavich, {\em {A Fake Split
  Supersymmetry Model for the 126 GeV Higgs}}.
  \href{http://dx.doi.org/10.1007/JHEP05(2014)113}{JHEP {\bf 05} (2014)  113},
  \href{http://arxiv.org/abs/1312.5220}{{\tt arXiv:1312.5220 [hep-ph]}}.

\bibitem{Braathen:2018htl}
J.~Braathen, M.~D. Goodsell, and P.~Slavich, {\em {Matching renormalisable
  couplings: simple schemes and a plot}}.
  \href{http://dx.doi.org/10.1140/epjc/s10052-019-7093-9}{Eur. Phys. J. C {\bf
  79} (2019) no.~8, 669}, \href{http://arxiv.org/abs/1810.09388}{{\tt
  arXiv:1810.09388 [hep-ph]}}.

\bibitem{Escudero:2005hk}
N.~Escudero, C.~Munoz, and A.~M. Teixeira, {\em {FCNCs in supersymmetric
  multi-Higgs doublet models}}.
  \href{http://dx.doi.org/10.1103/PhysRevD.73.055015}{Phys. Rev. D {\bf 73}
  (2006)  055015}, \href{http://arxiv.org/abs/hep-ph/0512046}{{\tt
  arXiv:hep-ph/0512046}}.

\bibitem{Kniehl:2016enc}
B.~A. Kniehl, A.~F. Pikelner, and O.~L. Veretin, {\em {mr: a C++ library for
  the matching and running of the Standard Model parameters}}.
  \href{http://dx.doi.org/10.1016/j.cpc.2016.04.017}{Comput. Phys. Commun. {\bf
  206} (2016)  84--96}, \href{http://arxiv.org/abs/1601.08143}{{\tt
  arXiv:1601.08143 [hep-ph]}}.

\bibitem{Kniehl:2015nwa}
B.~A. Kniehl, A.~F. Pikelner, and O.~L. Veretin, {\em {Two-loop electroweak
  threshold corrections in the Standard Model}}.
  \href{http://dx.doi.org/10.1016/j.nuclphysb.2015.04.010}{Nucl. Phys. B {\bf
  896} (2015)  19--51}, \href{http://arxiv.org/abs/1503.02138}{{\tt
  arXiv:1503.02138 [hep-ph]}}.

\bibitem{ParticleDataGroup:2020ssz}
{\bf Particle Data Group} Collaboration, P.~A. Zyla {\em et al.}, {\em {Review
  of Particle Physics}}. \href{http://dx.doi.org/10.1093/ptep/ptaa104}{PTEP
  {\bf 2020} (2020) no.~8, 083C01}.

\bibitem{Hempfling:1994ar}
R.~Hempfling and B.~A. Kniehl, {\em {On the relation between the fermion pole
  mass and MS Yukawa coupling in the standard model}}.
  \href{http://dx.doi.org/10.1103/PhysRevD.51.1386}{Phys. Rev. D {\bf 51}
  (1995)  1386--1394}, \href{http://arxiv.org/abs/hep-ph/9408313}{{\tt
  arXiv:hep-ph/9408313}}.

\bibitem{Degrassi:2007kj}
G.~Degrassi, P.~Gambino, and P.~Slavich, {\em {SusyBSG: A Fortran code for BR[B
  ---\ensuremath{>} X(s) gamma] in the MSSM with Minimal Flavor Violation}}.
  \href{http://dx.doi.org/10.1016/j.cpc.2008.06.012}{Comput. Phys. Commun. {\bf
  179} (2008)  759--771}, \href{http://arxiv.org/abs/0712.3265}{{\tt
  arXiv:0712.3265 [hep-ph]}}.

\bibitem{Bednyakov:2012rb}
A.~V. Bednyakov, A.~F. Pikelner, and V.~N. Velizhanin, {\em {Anomalous
  dimensions of gauge fields and gauge coupling beta-functions in the Standard
  Model at three loops}}. \href{http://dx.doi.org/10.1007/JHEP01(2013)017}{JHEP
  {\bf 01} (2013)  017}, \href{http://arxiv.org/abs/1210.6873}{{\tt
  arXiv:1210.6873 [hep-ph]}}.

\bibitem{Bednyakov:2012en}
A.~V. Bednyakov, A.~F. Pikelner, and V.~N. Velizhanin, {\em {Yukawa coupling
  beta-functions in the Standard Model at three loops}}.
  \href{http://dx.doi.org/10.1016/j.physletb.2013.04.038}{Phys. Lett. B {\bf
  722} (2013)  336--340}, \href{http://arxiv.org/abs/1212.6829}{{\tt
  arXiv:1212.6829 [hep-ph]}}.

\bibitem{Bednyakov:2013eba}
A.~V. Bednyakov, A.~F. Pikelner, and V.~N. Velizhanin, {\em {Higgs
  self-coupling beta-function in the Standard Model at three loops}}.
  \href{http://dx.doi.org/10.1016/j.nuclphysb.2013.07.015}{Nucl. Phys. B {\bf
  875} (2013)  552--565}, \href{http://arxiv.org/abs/1303.4364}{{\tt
  arXiv:1303.4364 [hep-ph]}}.

\bibitem{Luo:2002ey}
M.-x. Luo and Y.~Xiao, {\em {Two loop renormalization group equations in the
  standard model}}.
  \href{http://dx.doi.org/10.1103/PhysRevLett.90.011601}{Phys. Rev. Lett. {\bf
  90} (2003)  011601}, \href{http://arxiv.org/abs/hep-ph/0207271}{{\tt
  arXiv:hep-ph/0207271}}.

\bibitem{Bagnaschi:2014rsa}
E.~Bagnaschi, G.~F. Giudice, P.~Slavich, and A.~Strumia, {\em {Higgs Mass and
  Unnatural Supersymmetry}}.
  \href{http://dx.doi.org/10.1007/JHEP09(2014)092}{JHEP {\bf 09} (2014)  092},
  \href{http://arxiv.org/abs/1407.4081}{{\tt arXiv:1407.4081 [hep-ph]}}.

\bibitem{Kwasnitza:2020wli}
T.~Kwasnitza, D.~St\"ockinger, and A.~Voigt, {\em {Improved MSSM Higgs mass
  calculation using the 3-loop FlexibleEFTHiggs approach including
  $x_{t}$-resummation}}. \href{http://dx.doi.org/10.1007/JHEP07(2020)197}{JHEP
  {\bf 07} (2020) no.~07, 197}, \href{http://arxiv.org/abs/2003.04639}{{\tt
  arXiv:2003.04639 [hep-ph]}}.

\bibitem{Kwasnitza:2021idg}
T.~Kwasnitza and D.~St\"ockinger, {\em {Resummation of terms enhanced by
  trilinear squark-Higgs couplings in the MSSM}}.
  \href{http://dx.doi.org/10.1007/JHEP08(2021)070}{JHEP {\bf 08} (2021)  070},
  \href{http://arxiv.org/abs/2103.08616}{{\tt arXiv:2103.08616 [hep-ph]}}.

\bibitem{Allanach:2009ne}
B.~C. Allanach, G.~Hiller, D.~R.~T. Jones, and P.~Slavich, {\em {Flavour
  Violation in Anomaly Mediated Supersymmetry Breaking}}.
  \href{http://dx.doi.org/10.1088/1126-6708/2009/04/088}{JHEP {\bf 04} (2009)
  088}, \href{http://arxiv.org/abs/0902.4880}{{\tt arXiv:0902.4880 [hep-ph]}}.

\bibitem{Bagnaschi:2017xid}
E.~Bagnaschi, J.~Pardo~Vega, and P.~Slavich, {\em {Improved determination of
  the Higgs mass in the MSSM with heavy superpartners}}.
  \href{http://dx.doi.org/10.1140/epjc/s10052-017-4885-7}{Eur. Phys. J. C {\bf
  77} (2017) no.~5, 334}, \href{http://arxiv.org/abs/1703.08166}{{\tt
  arXiv:1703.08166 [hep-ph]}}.

\bibitem{Falk:1994es}
T.~Falk, K.~A. Olive, and M.~Srednicki, {\em {Heavy sneutrinos as dark
  matter}}. \href{http://dx.doi.org/10.1016/0370-2693(94)90639-4}{Phys. Lett. B
  {\bf 339} (1994)  248--251}, \href{http://arxiv.org/abs/hep-ph/9409270}{{\tt
  arXiv:hep-ph/9409270}}.

\bibitem{Arina:2007tm}
C.~Arina and N.~Fornengo, {\em {Sneutrino cold dark matter, a new analysis:
  Relic abundance and detection rates}}.
  \href{http://dx.doi.org/10.1088/1126-6708/2007/11/029}{JHEP {\bf 11} (2007)
  029}, \href{http://arxiv.org/abs/0709.4477}{{\tt arXiv:0709.4477 [hep-ph]}}.

\bibitem{Badziak:2014kea}
M.~Badziak, Z.~Lalak, M.~Lewicki, M.~Olechowski, and S.~Pokorski, {\em {Upper
  bounds on sparticle masses from muon g \ensuremath{-} 2 and the Higgs mass
  and the complementarity of future colliders}}.
  \href{http://dx.doi.org/10.1007/JHEP03(2015)003}{JHEP {\bf 03} (2015)  003},
  \href{http://arxiv.org/abs/1411.1450}{{\tt arXiv:1411.1450 [hep-ph]}}.

\end{thebibliography}\endgroup

\end{document}